\documentclass[twocolumn]{aastex631}

\usepackage[utf8]{inputenc}
\usepackage{amsmath,bm}
\usepackage{mathtools}   
\usepackage{graphicx}
\usepackage{xcolor}
\usepackage{verbatim}
\usepackage{comment}
\usepackage{soul}
\usepackage{changepage}
\usepackage[normalem]{ulem}

\newcommand*{\rom}[1]{\expandafter\@slowromancap\romannumeral #1@}

\begin{document}

\title{
        \Large
Modeling of lightcurves from reconnection-powered very high energy flares from M87*
        }

\author[0000-0001-6541-734X]{Siddhant Solanki}
\affiliation{Department of Astronomy, University of Maryland, 7901 Regents Drive, College Park, MD 20742, USA}

\author[0000-0002-2685-2434]{Jordy Davelaar}
\affiliation{Department of Astrophysical Sciences, Peyton Hall, Princeton University, Princeton, NJ 08544, USA}
\affiliation{NASA Hubble Fellowship Program, Einstein Fellow}

\author[0000-0002-7301-3908]{Bart Ripperda}
\affiliation{Canadian Institute for Theoretical Astrophysics, 60 St. George Street, Toronto, Ontario M5S 3H8}
\affiliation{Department of Physics, University of Toronto, 60 St. George Street, Toronto, ON M5S 1A7}
\affiliation{David A. Dunlap Department of Astronomy, University of Toronto, 50 St. George Street, Toronto, ON M5S 3H4}
\affiliation{Perimeter Institute for Theoretical Physics, 31 Caroline St. North, Waterloo, ON, Canada N2L 2Y5}

\author{Alexander Philippov}
\affiliation{Department of Physics, University of Maryland, 7901 Regents Drive, College Park, MD 20742, USA}

\begin{abstract} \label{abstract}
The black hole at the center of M87 is observed to flare regularly in the very high energy (VHE) band, with photon energies $\gtrsim 100$ GeV. The rapid variability, which can be as short as $2$ days in the VHE lightcurve constrains some of the flares to originate close to the black hole. Magnetic reconnection is a promising candidate for explaining the flares, where the VHE emission comes from background soft photons that Inverse Compton (IC) scatter off of high energy electron-positron pairs in the reconnecting current sheet. In this work, we ray trace photons from a current sheet near the black hole event horizon during a flux eruption in a magnetically arrested state in a general relativistic magnetohydrodynamics simulation. We incorporate beaming of the Compton up-scattered photons, based on results from radiative kinetic simulations of relativistic reconnection. We then construct VHE lightcurves that account for the dynamics of the current sheet and lensing from general-relativistic effects. We find that most of the flux originates in the inner $5$ gravitational radii, and beaming is essential to explain the observed flux from the strongest VHE flares. The ray traced lightcurves show features resulting from the changing volume of the reconnecting current sheet on timescales that can be consistent with observations. Furthermore, we find that the amount of beaming depends strongly on two effects: the current sheet inclination with respect to the observer and the anisotropy in the direction of motion of the accelerated particles.
\end{abstract}

\section{Introduction} \label{Introduction}

The elliptical galaxy Messier 87 (M87) is located at a distance of $16.8$ Mpc from the Milky Way \citep{M87_dist_2009ApJ...694..556B}. It hosts a supermassive black hole called M87*, with a mass of $6.5 \times 10^9 \text{M}_{\odot}$ \citep{M87_mass_2011ApJ...729..119G,EHT_1_2019ApJ...875L...1E}, where $\text{M}_{\odot}$ is a solar mass. M87* and a few other low luminosity active galactic nuclei (AGN) such as Sagittarius A* (Sgr A*), the supermassive black hole in our own Milky Way, show rapid variability, ranging from X-rays to, in the case of M87*, a few TeV $\gamma$-rays \citep{Sgr_Xray_2006A&A...450..535E,2012ApJ...746..151A}.
Multiple VHE flares have been observed from M87* in 2005, 2008, 2010 and 2018 \citep{Aharonian_2006,Albert_2008,Acciari_2010,2012ApJ...746..151A,EHT_broadband_2024arXiv240417623T}. These flares last for timescales as short as $\lesssim 2$ days, which is of the order of a few event horizon light-crossing times (about $8$ hours for M87*, assuming that it is rapidly spinning), and have fractional variability (i.e., the ratio of the fluxes during the flare and quiescent state, respectively) $\Delta F/F \sim 2-40$ \citep{EHT_broadband_2024arXiv240417623T}. The observed flux from the VHE flares is between $\sim 10^{-13} - 10^{-11}$ photons cm$^{-2}$ s$^{-1}$ for energies in the range $350 \text{ GeV}-10$ TeV. The implied isotropic-equivalent luminosities are in the range of $10^{40} - 10^{42}$ erg s$^{-1}$  \citep{Aharonian_2006,EHT_broadband_2024arXiv240417623T}\footnote{The estimate for the luminosity can vary by a factor of $\sim 2$, depending on the cutoff energy at which the spectrum becomes steeper.}. Furthermore, the correlated radio and X-ray emission from the M87* core in, e.g., the 2008 flare, points towards the near-horizon region as the origin of the flares, as opposed to a knot in the jet \citep{Acciari_2010,2012ApJ...746..151A}, where the latter was X-ray quiet \citep{2012ApJ...746..151A}. This is further supported by X-ray flares being detected from the M87* core \citep{chen_2023RAA....23f5018C}. Several mechanisms have been proposed to power the VHE flares, including curvature radiation from pairs in charge-starved regions inside the jet, gaps \citep{Levinson_PRL_2000PhRvL..85..912L,PIC_gap_2018A&A...616A.184L,PIC_gap_2018ApJ...863L..31C,Crinquand2020PRL}, or inverse Compton (IC) upscattering on magnetic reconnection-accelerated pairs from near the event horizon \citep{2020ApJ...900..100R,chashkina_2021MNRAS.508.1241C,2022ApJ...924L..32R,Hayk_2023ApJ...943L..29H}. Both mechanisms, which can potentially explain the energetics and variability of flares, require significant re-arrangement in the large-scale geometry and properties of the accretion flow during the flare. In the reconnection scenario, the formation of extended thin current sheets is required. In contrast, in the gap scenario, significant changes in the properties of the soft photon field are needed \citep{spark_gaps_response_2022ApJ...924...28K}. 

The plasma in the accretion disk around the black hole is nearly collisionless, with a 
luminosity of $L \sim 10^{-3}$ $L_{\text{{\rm Edd}}}$ is in units of the Eddington luminosity. The  Eddington luminosity is $L_{\text{{\rm Edd}}} \equiv {4 \pi G M m_p c}/{\sigma_T}$, where $G, M,  m_p, c$ and $\sigma_T$ are the gravitational constant, mass of black hole, mass of a proton, speed of light and the Thompson cross-section, respectively. The magnetic field strength of the black hole is estimated to be around $100$ G. This is inferred by matching the sub-millimeter radio flux observed by the Event Horizon Telescope (EHT) to synchrotron radiation from an accretion disk with a plasma $\beta = 1$ and a $1/r$ dependence of the magnetic field \citep{EHT_5_2019ApJ...875L...5E,EHT_2024A&A...681A..79E}, where $r$ is the distance from the black hole. Furthermore, this magnetic field strength agrees well with a jet powered by the Blandford-Znajek (BZ, \citep{1977MNRAS.179..433B}) mechanism \citep{2022ApJ...924L..32R}. Moreover, the radio and multiwavelength observations by EHT \citep{EHT_1_2019ApJ...875L...1E,EHT_5_2019ApJ...875L...5E,EHT8_2021ApJ...910L..13E,EHT_broadband_2021ApJ...911L..11E,EHT_near_horizon_2023ApJ...957L..20E,EHT_persistent_shadow_2024A&A...681A..79E,EHT_broadband_2024arXiv240417623T}, in particular of strong linear polarization, point towards the disk around the black hole to be in a magnetically-arrested state \citep{1974Ap&SS..28...45B,2003PASJ...55L..69N,2008ApJ...677..317I}. In this scenario, the pressure of strong ordered magnetic fields eventually becomes comparable to the ram pressure of the accreting plasma. General relativistic magnetohydrodynamics (GRMHD) simulations show that such disks exhibit quasi-periodic magnetic flux eruptions \citep{Tchekhovskoy_2011}, forming a large-scale current sheet where the magnetic fields at the base of the jet then reconnect \citep{2022ApJ...924L..32R}. The current sheet formed during these reconnection episodes is capable of accelerating electron-positron pairs to very high Lorentz factors due to the high magnetization expected in the base of the jet \citep{Hayk_2023ApJ...943L..29H}. 

One of the challenges of the reconnection scenario is that the duration of the reconnection event, as observed in the GRMHD simulations \citep{2022ApJ...924L..32R,Alisa_2023PhRvL.130k5201G}, is a factor of $10-100$ longer than the duration of the observed high-energy flares \citep{2012ApJ...746..151A}. The reconnection duration can be shorter in global kinetic simulations of accreting collisionless plasma, which have a reconnection rate that is faster by a factor of few \citep{Bransgrove_2021,Alisa_2023PhRvL.130k5201G}. While such kinetic simulations can account for particle acceleration in the presence of synchrotron and inverse-Compton cooling in a first-principles manner \citep[see, e.g.,][for local simulations]{wernerphilippov2019MNRAS.482L..60W,sironibeloborodov2020ApJ...899...52S,Hayk_2023ApJ...943L..29H,chernoglazov2023highenergy}, they can not yet capture the long-term steady-state of magnetically arrested flows and, thus, the three-dimensional global motion of the current sheet due to the computational expenses. Both of these effects can be very important in characterizing the lightcurves of high-energy flares. These effects can cause the radiation to get beamed in specific directions and potentially yield faster variability. For example, observations show that relativistic beaming near the black hole may be important in determining the rise and decay timescales of the flares \citep{von_Fellenberg_2023}.

In this work, we construct flare lightcurves by determining photon trajectories in global three-dimensional GRMHD data of a rapidly spinning black hole, capturing the near-horizon reconnection layer \citep{2022ApJ...924L..32R}. We use a novel Monte-Carlo (MC) scheme based on existing ray tracing codes \textsc{grmonty} and $\kappa$\textsc{monty} \citep{GRMONTY_2009ApJS..184..387D,kmonty_2023MNRAS.526.5326D}. We construct lightcurves relevant to the GeV-TeV band by tracing photons initialized close to the current sheet, with beaming prescriptions motivated by three-dimensional radiative particle-in-cell (PIC) simulations \citep{chernoglazov2023highenergy}. These simulations self-consistently include strong synchrotron cooling in a highly magnetized pair plasma, in the regime appropriate for low-luminosity AGN, such as Sgr A* and M87* . Our method of constructing the VHE lightcurves by introducing PIC-motivated beaming prescriptions into the GRMHD background can be applied to other relativistic astrophysical systems with large-scale current sheets (e.g., pulsar magnetospheres \cite{2016MNRAS.457.2401C}, neutron star collapse or merger remnants \cite{2024ApJ...968L..10S, 2024arXiv240401456M}, or black hole coronae \cite{2022ApJ...935L...1L}).

This paper is structured as follows. Section \ref{sec:VHE-IC Emission} contains the theory of VHE emission from the current sheet. Section \ref{Methods} describes the current sheet identification and ray tracing methods we adopted to construct the lightcurves. Section \ref{Results} contains the key results from our analysis. Section \ref{Discussion} discusses our results in context of VHE observations from low luminosity AGN and conclusions are presented in Section \ref{sec:Conclusions}.
 
\section{Reconnection-Powered Flares}\label{sec:VHE-IC Emission}
In the following section, we summarize the physics involved in the reconnection-powered VHE flare model. The process involves the following steps: (1) accretion of plasma onto the black hole, accompanied by increasing magnetic flux on the event horizon, and subsequent triggering of the large-scale magnetic reconnection; (2) acceleration of electron-positron pairs produced in the jet to large Lorentz factors inside the current sheet; and, (3) IC scattering of soft background photons by the energetic pairs to power VHE flares. We begin by listing the key physical parameters for M87* flares. 

\subsection{Physical Parameters}\label{VHE-IC:Physical Pars}
The inferred mass of the black hole and the local magnetic field strength from EHT are $M_{\text{BH}} = 6.5\times10^9 M_{\odot}$ and $B \sim 100$ G near the event horizon, respectively (in agreement with B $\sim 30$G at 7 gravitational radii ($r_{\rm g} = \frac{GM}{c^2}$) for a field linearly decaying with radius, as inferred by \cite{EHT_1_2019ApJ...875L...1E}). The jet power is of the order of $L_{{\rm jet}} \sim 10^{42} - 10^{44}$ ergs s$^{-1}$ \citep{10.1093/mnras/stw166}. Assuming that the jet is launched by the BZ mechanism, the theoretical jet power estimated using the inferred $100$ G magnetic fields matches the observed jet power. The $300$ GHz ($\sim 1$ meV photons) flux near the M87* horizon is $\sim 1$ Jy, from which one can estimate the soft photon background energy density at $1$ meV to be $U_{\mathrm{soft}} \approx 0.01$ ergs cm$^{-3}$ \citep{Broderick_S_2015ApJ...809...97B,EHT_broadband_2021ApJ...911L..11E}. This energy density is representative of the quiescent state of the accretion disk, and it can be a factor of a few lower during the magnetic flux eruption event when the accretion disk, the main source of the meV emission, is ejected from near the black hole \citep{2022ApJ...924L..32R,Jia_2023arXiv230109014J}. The accretion rate onto the black hole is $\sim 10^{-3} M_{\odot} \text{yr}^{-1}$ \citep{EHT_1_2019ApJ...875L...1E}.

Two other important parameters affect the microphysics of particle acceleration: the radiation reaction limited Lorentz factor and the plasma magnetization. The radiation reaction-limited Lorentz factor, $\gamma_{{\rm rad}}$, is the Lorentz factor at which the radiative losses balance out the acceleration of pairs due to the electric field in the current sheet. Since the magnetic energy density vastly exceeds the energy density of the soft radiation field, $B^2/8\pi \gg U_{\rm soft}$, radiation reaction is dominated by synchrotron losses. Additionally, the plasma magnetization, $\sigma_c = b^{2}/ (4 \pi \rho_{e^{\pm}} c^2)$ is the ratio of twice the magnetic energy density in a frame co-moving with the fluid (${b^2}/{4\pi} = {b^{\mu} b_{\mu}}/{4\pi}$) to the rest mass energy of the pairs, where $\rho_{e^{\pm}}$ is the mass density of pairs in the current sheet. Here, $b^{\mu}$ is the magnetic field 4-vector; refer to \cite{Noble_2006}
for its definition and relation to $B$ (note that we use Greek indices to represent $4$-vectors and Latin indices to represent $3$-vectors throughout the paper). 
The characteristic energy gain of particles accelerated during reconnection is set by this magnetization, $\gamma \sim \sigma_c$. Note that most of the pair density in this scenario is produced locally, by collisions of high-energy synchrotron photons produced by reconnection-accelerated particles. We follow the estimates of \cite{Hayk_2023ApJ...943L..29H} for $\gamma_{{\rm rad}}$ and $\sigma_c$ to be $4 \times 10^6$ and $\sim 10^7$, respectively. The value of $\sigma_c$ depends on the plasma number density, which is assumed to be set by the pairs produced from the synchrotron radiation from the energized particles in the current sheet \citep{Hayk_2023ApJ...943L..29H}.

Three-dimensional PIC simulations that account for radiative losses have shown how particle acceleration happens in both weak, ${\gamma_{{\rm rad}}}/{\sigma_c} > 1$, and strong, ${\gamma_{{\rm rad}}}/{\sigma_c} < 1$, cooling regimes \citep{chernoglazov2023highenergy}. 
In the weak cooling regime, the highest energy particles, $\gamma \sim \gamma_{\rm rad}$ are accelerated by the reconnection-driven electric field while bouncing between the two converging upstreams \citep{2021ApJ...922..261Z}.
In the strong cooling regime, however, the most energetic particles, $\gamma \sim \sigma_c$, preferentially move along the upstream magnetic field, escaping the synchrotron burnoff limit. When $\gamma_{{\rm rad}} \sim \sigma_c$, the cooling is moderate and the pairs' velocities at highest energies are almost isotropic. Following \cite{Hayk_2023ApJ...943L..29H}, we generally expect ${\gamma_{\rm rad}}/{\sigma_c} \lesssim 1$ for M87*, putting it into the moderate to strong cooling regime. In this paper we construct lightcurves for both isotropic and along-the-upstream field beaming prescriptions to model both the moderate and strong cooling regimes. {Additionally, an isotropic beaming prescription can be applicable if the magnetic field upstream of the current sheet is much less ordered on microscopic scales than what GRMHD simulations performed at macroscopic scales show. This situation can be realized if vigorous turbulence is excited during the current sheet formation.}

\subsection{VHE Flares: Energy Source}\label{VHE-IC:particle_energization}
In order to produce TeV energy photons, electrons and positrons need to have energies of at least $1$ TeV or Lorentz factors greater than $\gamma = E / (m_e c^2) \gtrsim 2 \times 10^6$, where $E$ is their energy. The energy to accelerate the pairs during reconnection comes from the dissipation of the magnetic field energy, which is the free energy in the system. Accelerated pairs of $\sim 1$ TeV energies are cooled extremely quickly, $t_{{\rm cool}} \sim 1 \text{ TeV} / P_{{\rm sync}} \approx 0.1 \text{ s}$, where $ P_{{\rm sync}}=(4/3) \sigma_T c \gamma^2 (B^2 / 8\pi)$ is the synchrotron power. The almost instantaneous nature of the synchrotron cooling allows us to not track accelerated particles through GRMHD scales but instead postulate instantaneous ``sub-grid'' emissivity (see Sec. \ref{subsec: photon initialization}). 

The synchrotron radiation has a characteristic luminosity of $L_{{\rm rec}} \approx 0.1 L_{{\rm jet}}$, where $L_{jet}$ is the M87* jet luminosity \citep{Hayk_2023ApJ...943L..29H}. The factor of $0.1$ comes from the amount of magnetic flux that is reconnected, which energizes the pairs that then radiate away their energy via synchrotron emission. The VHE photons are produced by the up-scattering of the soft photons from the disk by particles accelerated in the current sheet. The ratio of synchrotron to Inverse-Compton luminosity can be approximated as $L_{\rm rec} / L_{\rm IC} \sim \Bigl(<\gamma^2 B_{\perp}^2>/8\pi \Bigr) / \Bigl(<\gamma^2> U_{\rm soft} \Bigr) \sim \Bigl(\gamma_{\rm rad} / \sigma\Bigr)^2 U_{\rm B} / U_{\rm soft} \sim 4\times 10^{2} \text{ } (\gamma_{\rm rad}/\sigma)_{0.1}^{2} B_{100}^2 U_{\rm soft, 0.01}$, where the subscripts represent the values by which the quantities are normalized. Here, $<\cdots>$ denotes an average over the particle distribution function, $\gamma$ is the Lorentz factor of the particles, $B_{\perp}$ is the effective perpendicular field that the particles experience, and $U_B$ is the magnetic field energy density. Simulations of \cite{Hayk_2023ApJ...943L..29H} show that $<\gamma^2 B_{\perp}^2> / 8\pi \sim \gamma_{rad}^2 U_B$ and $<\gamma^2> \sim \sigma^2$. For a maximally spinning black hole, the jet luminosity can be estimated as $L_{\rm jet} \sim c B^2 r_g^2 \sim 2 \times 10^{44} \text{ B}^2_{100} \text{ erg s}^{-1}$. This yields a total IC luminosity of $\sim 6 \times 10^{40} \text{ erg s}^{-1}$ \citep{Hayk_2023ApJ...943L..29H}. The isotropic equivalent luminosity inferred from the strongest VHE flares of M87 can reach $10^{42} \text{ erg s}^{-1}$ \citep{2012ApJ...746..151A,EHT_broadband_2024arXiv240417623T}, which requires some amount of beaming to be present in the reconnection-powered flare model.
\newline
Beaming effects will also be present in the synchrotron radiation emitted by energetic particles in the current sheet. In particular, simulations by \cite{chernoglazov2023highenergy} show that the highest energy particles in the reconnecting current sheet produce radiation in excess of the synchrotron burnoff limit, $\sim 16 {\rm MeV}$, which is strongly beamed along the upstream magnetic field, i.e., in the same direction as the VHE IC emission that we consider in this paper.


\subsection{VHE Flares: Angular Distribution of Power}\label{VHE-IC:angular_distribution_of_power}
While previous works have shown that IC emission by reconnection-accelerated particles is a plausible mechanism to get the correct VHE flux luminosity, they do not study the variability and angular distribution of the IC emission, which are important to understand the lightcurves and compare the model to observed data. Potential sources of variability and beaming include (a) local anisotropy in the distribution function of accelerated pairs (that is, accelerated particles preferentially moving along or perpendicular to the background magnetic field), (b) the changing volume and geometry of the current sheet, (c) the variation in the local IC emissivity throughout the current sheet, and (d) general relativistic effects in the vicinity of the black hole\footnote{In addition to these effects, the faster flux decay rate in collisionless plasma, compared to the GRMHD approximation, may also contribute to faster variability due to the flux eruptions \citep{Bransgrove_2021,Alisa_2023PhRvL.130k5201G}.}.

As discussed previously, VHE flares can potentially be powered by the IC emission of pairs accelerated in the current sheet. In the strong cooling regime, $\gamma_{\rm rad}/{\sigma_c} < 1$, pairs that manage to exceed the synchrotron burnoff limit preferentially move along the upstream magnetic field, $\hat{B}_{up}$ \citep{chernoglazov2023highenergy}.
Thus, the IC radiation powered by the energetic pairs can be highly beamed along the direction of the local upstream magnetic field. The global magnetic field direction plays an important role in determining how much of the flare luminosity is received by a distant observer. The local beaming effects can be washed out if the magnetic field around the current sheet is randomly oriented. On the other hand, an ordered magnetic field will preserve some of local beaming and make the overall IC flux anisotropic. The coherent near-horizon sub-millimeter polarization from M87* points towards ordered magnetic fields around the black hole, where this effect can be important \citep{EHT8_2021ApJ...910L..13E}. Moreover, the current sheet during the flux eruption itself is dynamic. The changing volume and geometry of the current sheet can, in principle, affect the flare luminosity. 
In particular, if the volume of the current sheet changes over time then it will also manifest as a variability in the lightcurve.

Finally, since in the reconnection scenario, the magnetic field energy at the jet base powers the flares, the local VHE emissivity scales as the local magnetic field energy density. While the exact radial profile of the magnetic field is unknown for M87*, the field strength should generally scale as $\sim 1/r$ for a preferentially toroidal field \citep{2020ApJ...900..100R}, and as $\sim 1/r^2$ for a poloidal field. Consequently, the VHE emissivity is expected to drop off with distance from the black hole, following the fluid-frame magnetic energy density, $b^{\mu}b_{\mu} \sim B^2 \Gamma^{-2}$, where $\Gamma$ is the Lorentz factor of the plasma. Following these points, below we describe our procedure of constructing the VHE lightcurves from the GRMHD simulation using a sub-grid model for the beaming of highest energy particles, based on results of radiative PIC simulations \citep{chernoglazov2023highenergy,Hayk_2023ApJ...943L..29H}.

\section{Methods} \label{Methods}
We analyze a high-resolution GRMHD simulation of a spinning black hole (where the spin parameter $a = 0.9375$) with a geometrically thick accretion disk that reaches a magnetically arrested state, presented in \cite{2022ApJ...924L..32R}. The simulation is performed with the Graphics Processing Unit (GPU)-accelerated code \textsc{HAMR} \citep{HAMR_Liska_2022}. The simulation is performed in spherical Kerr-Schild coordinates ($t, r, \theta, \phi$) and has an (effective) resolution of $N_r \times N_{\theta} \times N_{\phi} = 5376 \times 2304 \times 2304$, and a radial domain of $r \in \{1.2, 2000\}r_{\rm{g}}$. It is initialized with a large-scale poloidal (in the $r-\theta$ plane) magnetic field to ensure that the simulation reaches the MAD state. 
The strength of the initial magnetic field is normalized such that $2 P_{max} / (b^{\mu}b_{\mu})_{max} = 100$, where $P$ is the gas pressure, and the `$\_{max}$' values are the maximum values in the entire grid. The adiabatic index of the gas is $\hat{\gamma} = 13/9$, which lies between the adiabatic indices of relativistic and non-relativistic gases, respectively. MHD simulations require lower bounds (floors) on fluid variables such as pressure and density.  In the simulation, the density floor of the plasma in the jet is set such that the maximum value of the plasma magnetization in the domain is $\sigma_{c,\mathrm{max}} = 25$, using the magnetic field strength in the frame co-moving with the fluid. The simulation is performed with a sufficiently high resolution so that the current sheet near the event horizon shows plasmoid formation, which indicates that the reconnection occurs in the fast asymptotic MHD regime. This observation is confirmed by the measured reconnection rate of $\sim 0.01 c$ \citep{2022ApJ...924L..32R}, where the reconnection rate is defined as the inflow velocity into the sheet, normalized by the Alfv\'{e}n speed, $v_{\rm A} = \sqrt{\sigma_{c} / (\sigma_{c} +1)} c \sim c$ for $\sigma_c = 25$ in the jet. 

The GRMHD outputs have a cadence of $5$ $t_{\text{g}}$ where $t_{\text{g}} = r_{\rm{g}}/c = GM/c^3$ in the Kerr-Schild (KS) coordinate time. The simulation shows three large-scale magnetic flux eruption events. Our analysis is done in the time interval $9050-9750$ $t_{\text{g}}$, which brackets one of the eruption events or the flaring state. A large-scale current sheet forms during the flaring state and the magnetic flux at the horizon rapidly drops \citep{2022ApJ...924L..32R}. 

The current sheet is confined within approximately $10$ $r_{\rm{g}}$, so we truncate our analysis domain within that radius. Similarly, the current sheet does not move more than $20^{\circ}$ from the equator, so we also limit the analysis domain to $\theta \in (45^{\circ}, 135^{\circ})$.

To construct the VHE lightcurves, we perform the following steps: (1) identify the current sheet -- location where pairs would be accelerated to large Lorentz factors, and the high energy emission would be produced; (2) initialize the wavevectors of the photons in accordance with the local magnetic field geometry, and the different prescriptions for the amount of beaming of accelerated particles; and (3) ray trace photons to a large distance from the black hole following the method based on the ray tracing codes \textsc{grmonty} and \textsc{$\kappa$monty } \citep{GRMONTY_2009ApJS..184..387D,kmonty_2023MNRAS.526.5326D}. We now describe each of the steps in more detail.

\subsection{Current Sheet Identification} \label{Subsection: Current Sheet Identification}
 
The current sheet in the simulation is identified using a combination of the cold magnetization parameter, $\sigma_{c}$, and dimensionless fluid temperature, $T = P / \rho$ (i.e., in units of ion rest mass energy), thresholds. We expect a small magnetization parameter, $\sigma_c$, in the current sheet as well as a high sheet temperature proportional to the upstream magnetization (i.e., the temperature depends on $\sigma_c$ in the upstream, and not on the small $\sigma_c$ inside the current sheet), $T \propto \sigma_{\rm{c,max}} > 1$ \citep{2022ApJ...924L..32R}. The value of $\sigma_c$ is small inside the current sheet because $b^{\mu}b_{\mu}$ drops in the reconnection region and the density peaks in the current sheet compared to the magnetized upstream\footnote{Reconnection in the current sheet present during the magnetic flux eruption occurs in the zero-guide field regime, see \cite{2022ApJ...924L..32R}.}. The fluid is heated up during reconnection as the (upstream) magnetic energy is converted into the thermal and kinetic energies of the plasma. The current sheet is more clearly captured using these two thresholds than directly computing the current density, which does not always uniquely identify reconnecting current sheets  (see also Figure 8 of \citealt{2020ApJ...900..100R}).

\begin{figure*}
    \centering
    \includegraphics[width=0.99\textwidth]{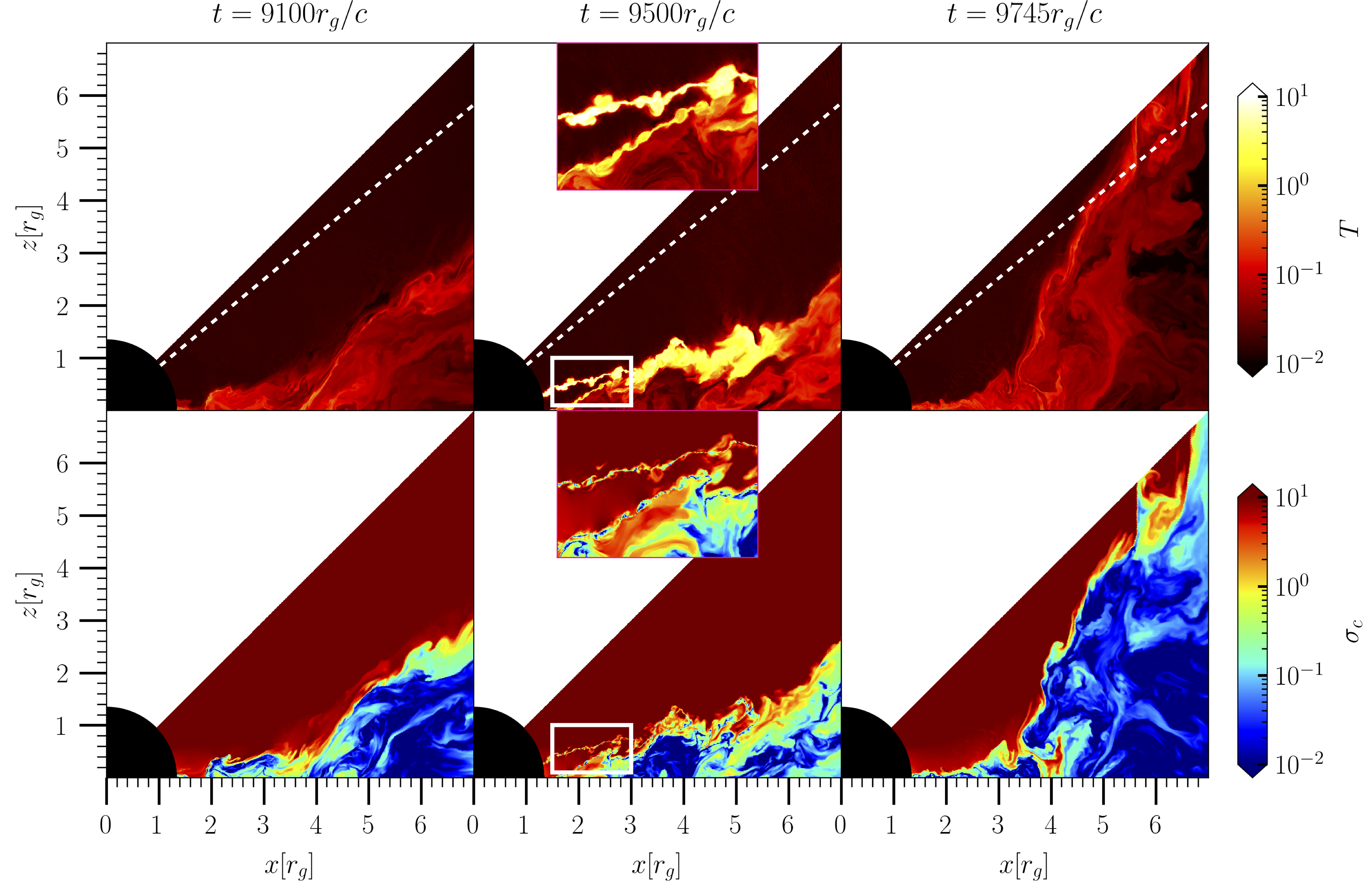}
    \caption{Poloidal slices of the dimensionless temperature, $T$, and magnetization, $\sigma_c$, taken during three snapshots in the simulation. The snapshots are taken before, during and after a large-scale magnetic flux eruption respectively. Fluid heats up to temperatures $T \sim \sigma_{\text{jet}}$ in the middle panel $(t=9500 r_{\rm{g}}/c)$ as a result of magnetic reconnection. Likewise, $\sigma_c$ is nearly zero where reconnection takes place because of vanishing magnetic fields and accumulation of matter in the current sheet. We use $T>1$ and $\sigma_c < 0.01$ as our thresholds for current sheet identification. The insets in the middle panels show a zoomed-in view of the current sheet that undergoes reconnection. The plasma in the sheet is characterized by low magnetization, $\sigma$, and high temperature, $T$. We calculate the upstream magnetic field strength $b^2=b^{\mu}b_{\mu}$ profiles in Figure \ref{fig:bsq_profile}, along the white dashed lines.}
    \label{fig:temperature_sigma_slice}
\end{figure*}

\begin{figure}
    \centering
    \includegraphics[width=0.49\textwidth]{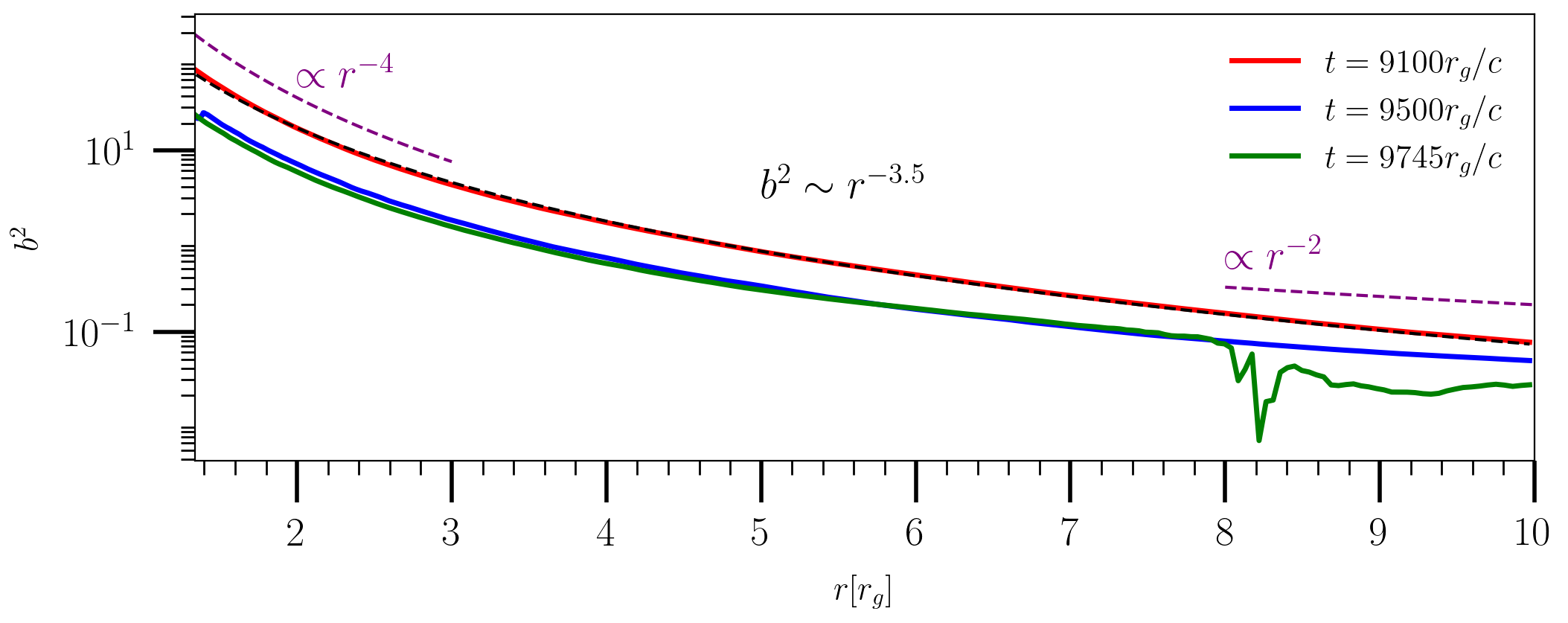}
    \caption{The radial profile of $b^2 = b^{\mu}b_{\mu}$, representing the fluid frame magnetic field energy, taken along the white dashed lines in Figure \ref{fig:temperature_sigma_slice}.  $b^2 \sim r^{-3.5}$ drops during the magnetic flux eruption, once the magnetic field lines start reconnecting. A similar drop in $b^2$ during a magnetic flux eruption is seen in Figure 14 of \cite{2020ApJ...900..100R}. The approximate $r^{-3.5}$ scaling can be explained by the field being mainly radial, $b^2 \propto r^{-4}$, near the black hole, and increasingly toroidal, $b^2 \propto r^{-2}$, at larger distances. } 
    \label{fig:bsq_profile}
\end{figure}

Figure \ref{fig:temperature_sigma_slice} shows the values of $T$ and $\sigma_c$ for three snapshots at different times in the simulation. The panels from left to right are taken before, during and after a large-scale magnetic flux eruption, respectively. One can see the formation of a thin current sheet in the zoomed-in insets in the middle panels $(t = 9500$ $r_{\rm{g}}/c)$, evident from the increase in fluid temperature and decrease in $\sigma_c$. The white dashed lines in the two figures are the locations where the fluid frame magnetic field energy profile, $b^2$, is computed and plotted in Figure \ref{fig:bsq_profile}. It is also clear that a combination of both parameters is necessary to pick out the current sheet. For instance, the middle panel of Figure \ref{fig:temperature_sigma_slice} contains high-temperature zones that are clearly not located in the current sheet (e.g., hot fluid advected out of the reconnection zones). Similarly, the bulk of the disk has a low value of the magnetization, $\sigma_c$, and is not a part of the current sheet.

 There is no radial dependence of $\sigma_c$ in the jet very close to the black hole, $\lesssim 10$ $r_{\rm{g}}$, and $\sigma_{c,\text{jet}} \approx \sigma_{\text{max}}$ because of the imposed density floor. As a result, the plasma temperature in the reconnection layer is independent of the distance from the black hole as well, which motivates us to use a constant temperature threshold for the reconnection zone. The temperature in the disk-jet boundary during the magnetic flux eruption is $\gtrsim 1$, much higher than the temperature everywhere else in the disk. We take this threshold as a good indicator of localizing the plasma that passed through the reconnection zone. Likewise, we adopt a constant $\sigma_c$ threshold of $\sigma_c < 0.01$ to locate the regions of weak magnetic field, that is, in the current sheet. We test various choices of $T$ and $\sigma_c$ thresholds in Appendix \ref{Appendix: thresholds for current sheet identification}, and find that the above thresholds adequately capture the location of the current sheet. To summarize, we use constant thresholds of $T > 1$ and $\sigma_c < 0.01$ to identify the current sheet.

\subsection{Photon Beaming and Upstream Identification} \label{Subsection: Photon Beaming and Upstream Identification}
After identifying the current sheet, we need to find the direction of the upstream magnetic field, $\hat{B}_{up}$, along which the accelerated particles can be beamed. The VHE photons are beamed along the direction of motion of the high-energy particles that do the scattering. As a result, in a strongly cooled regime, $\gamma_{{\rm rad}}\lesssim\sigma_c$, the IC photons are primarily emitted along the upstream magnetic field around the current sheet. The distribution of photon emission directions becomes more isotropic as $\sigma_c$ approaches $\gamma_{{\rm rad}}$ \citep{chernoglazov2023highenergy}. The results of PIC simulations directly apply to reconnection in a flat spacetime with no bulk flows in the upstream. However, there are global flows in the jet in the vicinity of the black hole, determined by the $\mathbf{E} \times \mathbf{B}$ drift resulting from the rotation of field lines, as well as possible bulk flows along the field lines. Because of these effects, we first identify the locally flat fluid rest frame in the upstream of the current sheet. We initialize the photon momenta in this frame according to the above prescriptions from PIC simulations, and boost back into the lab frame for raytracing\footnote{We verified that plasma in the upstream of the current sheet mainly moves according to the ${\bf E}\times {\bf B}$ drift, as expected for relativistic reconnection. Thus, in flat spacetime our prescription in the case of strong synchrotron cooling is equivalent to boosting photons along ${\bf \hat{v}}=\hat{v}_{\parallel}{\bf B}/B + {\bf E}\times {\bf B}/B^2$, where $\hat{v}_{\parallel}$ is chosen such that $|{\bf{\hat{v}}}|=1$. Such a prescription was used to model $\gamma$-ray lightcurves from pulsars \citep{Bai2010} and is consistent with global PIC simulations of pulsar magnetospheres \citep{2016MNRAS.457.2401C}}.   

The structure of turbulent $3$D magnetic field inside the identified current sheet does not allow to easily construct a vector along the upstream field. Instead, we consider a spherical shell around each cell in the previously identified current sheet. We then define the upstream field to be the magnetic field vector with the largest $b^2$ in the scanned volume, which is also above a threshold, $b^2 \geq 80 r^{-3.5}$. The numerical pre-factor is chosen such that the selected values correspond to the true upstream field, and not local field compressions inside the current sheet. 
The radial dependence is chosen such that we are not biased in picking the magnetic field directions corresponding to cells closer to the black hole, where the field strength is highest. Note that $b^2$ in the upstream of the sheet approximately scales as $\sim r^{-3.5}$ (see the $b^2$ profile along the jet in Figure \ref{fig:bsq_profile}).

\begin{figure}
    \centering
    \includegraphics[width=0.49\textwidth]{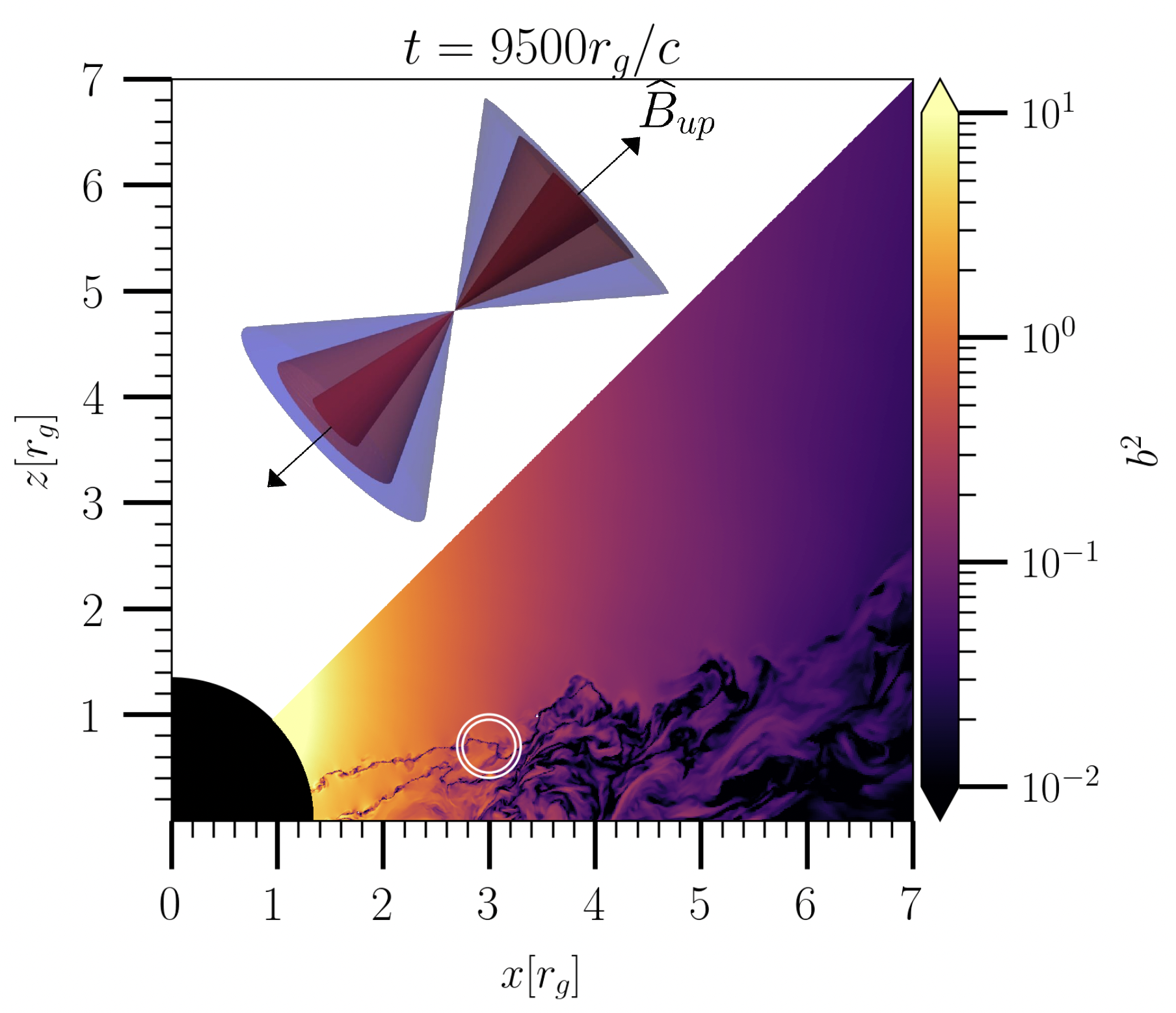}
    \caption{Polar slice of $b^2$ taken during the magnetic flux eruption. The current sheet is the region where $b^2$ falls as the magnetic field lines reconnect. In order to identify the upstream magnetic field, we take a small volume in the form of a thick spherical shell around each cell in the current sheet. The white shell plotted in the figure is representative of the volume we scan around every cell and pick the largest magnetic field scaled out by its radial dependence. Once the corresponding upstream magnetic field is identified for every point in the current sheet, the photons' (superphotons') wavevectors are sampled by using a Gaussian distribution centered around the upstream field. The cones on the top left represent different amounts of beaming, with smaller opening angles corresponding to stronger beaming.}
    \label{fig:bsqr_slice}
\end{figure}

We find that the maximum value of $b^2$ computed within spherical shells saturates at a distance of $\approx 0.2$ $r_{\rm{g}}$ away from the current sheet. Therefore, we select the inner and outer radii of the scanned shells to be $0.25$ and $0.3$  $r_{\rm{g}}$, respectively. A polar slice of the $b^2$ values is shown in Figure \ref{fig:bsqr_slice}, taken during the flux eruption at $t=9500$ $r_{\rm{g}}/c$.  We overplot a shell with inner and outer radii of $0.25$ and $0.3$ $r_{\rm{g}}$ respectively (white rings), representative of the volume scanned to find the upstream field. We find that the choice of the outer radius of the spherical shell does not change the overall distribution of the identified upstream field as long as it is between $0.3 - 0.4$ $r_{\rm{g}}$. This convergence test is discussed in more detail in Appendix (\ref{Appendix: selection of the upstream magnetic field}).

Once the direction of the upstream magnetic field is identified, $b^{\mu}$ is transformed to the fluid frame. This is done by constructing a tetrad basis, $\hat{e}_{\mu}$, where the time axis, $\hat{e}_{t}$, is along the four-velocity of the fluid in the upstream, and $\hat{e}_{x}$ points along the identified upstream magnetic field. The $\hat{e}_{y}$ and $\hat{e}_{z}$ are constructed using the Gram-Schmidt orthonormalization, so the tetrad axes are orthonormal in metric $g_{\alpha \beta}$. The fluid frame magnetic field is then obtained by projecting the lab frame magnetic field vector along the tetrad. 

\subsection{Photon Initialization and Geodesic Integration} \label{subsec: photon initialization}
We trace photons originating from the current sheet using a MC based method following \textsc{grmonty} and \textsc{$\kappa$monty} \citep{GRMONTY_2009ApJS..184..387D,kmonty_2023MNRAS.526.5326D}. We initialize and ray trace packets of photons called `superphotons'. With the current sheet and the upstream magnetic field identified, we initialize $N_i(\text{x}^{\mu}_i)$ superphotons with weights $w_i$ and wavevectors $\text{k}^{\mu}_i = (k^0_i, \vec{k}_i)$ in every cell $i$ inside the current sheet. Here, $\text{x}^{\mu}_i$ is the $4-$position of cell $i$ in Kerr-Schild coordinates. The superphotons are initialized in a frame that is co-moving with the upstream fluid and is centered on the current sheet cell $i$. The $x$ axis is defined to be along the upstream magnetic field, $\hat{B}_{{\rm up,i}}$, identified previously.

We use a random number generator (RNG) to sample $|k_{\perp, i}| = \sqrt{k_{y,i}^2 + k_{z,i}^2}$, the norm of the perpendicular component of the superphoton's wavevector with respect to the upstream magnetic field, using a Gaussian probability distribution. The distribution has a mean of $0$ and a standard deviation of $\Delta \chi$, which sets the opening angle of the photon beams coming out of the sheet. $\Delta \chi$ quantifies the amount of beaming: when it is small,  $\vec{k}_i$ is mostly along the local $\hat{B}_{{\rm up, i}}$, as it is expected in the case of strong cooling (accelerated pairs move along $\hat{B}_{up, i}$ and radiate in the same direction). Conversely, when $\Delta \chi$ is large, there is no correlation between $\vec{k}_i$ and $\hat{B}_{up, i}$,  representative of the case of weak cooling. Superphotons have $k^{\mu}k_{\mu} = 0$ and  $|k| = 1$, implying $|k_{\perp}| \leq 1$. However, the RNG can occasionally sample unphysical wavevectors, with $|k_{\perp}| > 1$ ($\implies k^{\mu}k_{\mu} \neq 0$). When this happens, we re-sample $|k_{\perp}|$ . To construct the full wavevector, we use uniform distributions to sample an azimuthal angle $\phi_i \in [0, 2\pi]$ and the sign of the $k_{\parallel, i}$, the wavevector component parallel to $\hat{B}_{up, i}$, $\mathrm{sign}(k_{\parallel, i}) \in \{-1, 1\}$. The latter is done to ensure that superphotons can propagate both parallel and anti-parallel to the upstream field. Finally, ${k}^{\mu}$ is initialized such that $k^0 > 0$ and ${k}^{\mu} {k}_{\mu} = 0$. We do not consider the energy distribution of the photons since all the $\vec{k}_i$'s have a unit norm. This is because the IC emission mechanism is treated using a prescription for beaming in post-processing, and no information about the distribution function of emitting pairs that produce the IC radiation is specified, or required. The different cones correspond to different values of $\Delta \chi$ in Figure \ref{fig:bsqr_slice}. In this work, we consider a few choices of $\Delta \chi$, namely $ \Delta \chi \in \{1^{\circ}, 10^{\circ}, 45^{\circ}\}$, going from strongly to weakly beamed photons along the magnetic field respectively.  We also include a model with no sub-grid beaming, where $k_x, k_y \text{ and } k_z$ are all sampled from a uniform distribution. 

The number of photons emitted in each cell, $N_i$, is proportional to the volume element as well as the local IC emissivity, i.e., $N_i \propto j(\text{x}_i^{\mu}) \text{d}^3\text{x}^{\mu}_i \sqrt{|-g(\text{x}^{\mu}_i)|}$, where $g(\text{x}^{\mu}_i)$ is the determinant of the metric. The coordinates used in the simulation have constant $\text{d}^3\text{x}^{\mu}_i$ and therefore $N_i \propto j(\text{x}_i^{\mu}) \sqrt{|-g(\text{x}^{\mu}_i)|}$.

Further, $j(\text{x}_i^{\mu})$ is the IC emissivity of the VHE photons, that we set to be proportional to $b^2(\text{x}_i^{\mu})$. This choice is motivated by the fact that the average energy density of the accelerated leptons in the current sheet is proportional to the upstream magnetic field energy density. In practice, $N_i$ can have very large values for different cells in the grid, making ray tracing more computationally expensive. To circumvent this issue, we define a quantity $c_i = {\sqrt{|-g(\text{x}^{\mu}_i)|}}/{\text{min}(\sqrt{|-g(\text{x}^{\mu})|})_{\forall i}}$ and set $N_i = c_i^{1/2}$ and $w_i =  b^2(\text{x}_i^{\mu}) c_i^{1/2}$. This ensures that at least $1$ photon is produced in every cell that belongs to the current sheet. The superphoton weight, $w_i$, depends on the upstream magnetic field strength and the determinant of the metric at the position of emission. We obtain identical results using $N_i = 1 \times c_i^{1/2}, 5 \times c_i^{1/2} \text{ and } 10 \times c_i^{1/2}$ (i.e., ray tracing at least $1, 5$ and $10$ superphotons per current sheet cell respectively), as shown in the Appendix 
\ref{Appendix: Convergence of the Lightcurve}. 

The $N_i$ wavevectors from every current sheet cell are then transformed back to the lab frame. We use the standard Runga-Kutta 4 (RK4) integrator to integrate the geodesics in modified (i.e., logarithmic in radius) Kerr-Schild coordinates. We use adaptive timestepping described in \cite{GRMONTY_2009ApJS..184..387D} with an initial stepsize of $10^{-4}$ $t_g$. A small stepsize is required because most of the emission originates very close to the black hole, where general-relativistic effects are important and smaller steps are required to maintain accuracy. The convergence of our ray tracing scheme is discussed in Appendix \ref{Appendix: Convergence of the Lightcurve}.  Numerical integration ends when the geodesics cross the event horizon or attain a coordinate radius $r > 10^4 r_{\rm{g}}$.  The lightcurves are constructed from binning the superphotons that reach the outer boundary, weighted by $w_i$. We analyze the total data of $140$ simulation outputs, spanning $t \in (9000$ $t_g, 9750$ $t_g)$, that capture a magnetic flux eruption that occurs during the interval $(9300$ $t_g, 9600$ $t_g)$.

\section{Results} \label{Results}
\subsection{Emission Zone}\label{Results: emitting regions}
\begin{figure*}
    \includegraphics[width=0.99\textwidth]{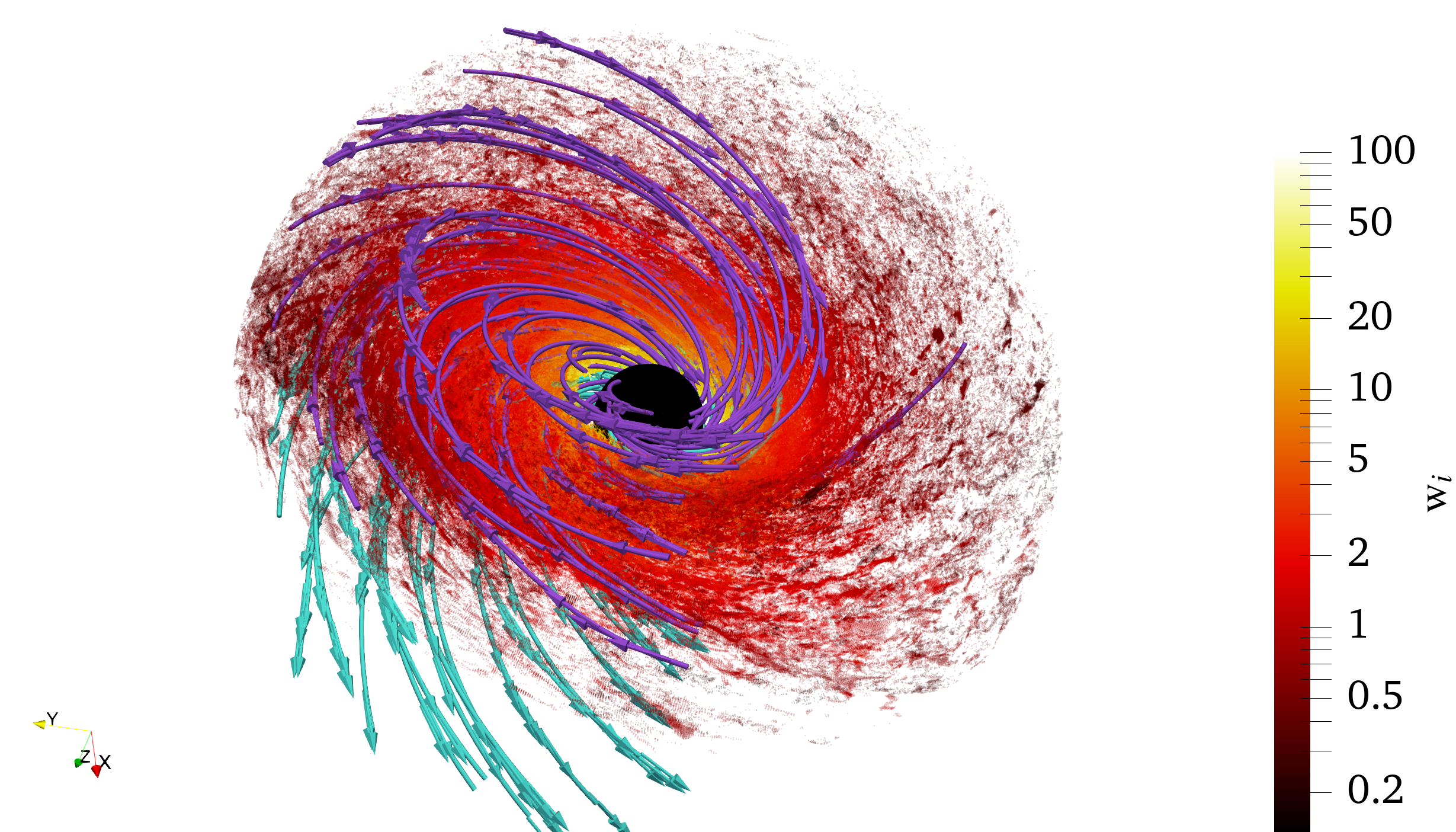}
    \caption{The location of the VHE emission production, combined over all of the simulation snapshots analyzed in this paper. The color represents the superphoton weights $w_i$. The emission region lies in the equatorial plane of the simulation at smaller distances but gets progressively misaligned from it further from the black hole. The magnetic field lines are over-plotted as purple (pointing radially inwards) and cyan (pointing outwards) lines, taken from one of the simulation snapshots during the magnetic flux eruption. }
    \label{fig:full_current_sheet}
\end{figure*}

The identified locations of VHE photon production, combined over all of the analyzed simulation snapshots, are shown in Figure \ref{fig:full_current_sheet}. The magnetic field streamlines are over-plotted, which are taken from a snapshot during the state of magnetic eruption at $t = 9500 r_{\mathrm{g}}/c$. The purple and cyan field lines represent field lines that point radially inwards and outwards, respectively\footnote{In the rest frame of the upstream plasma moving nearly radially, the wavevectors of superphotons are aligned along the upstream magnetic field in the strongly beamed models.}. The colors represent the superphoton weights $w_i \propto b^2(\text{x}^{\mu}_i) \sqrt{|-g(\text{x}^{\mu}_i)|}$ \footnote{This prescription assumes that the amount of the dissipated power scales with the magnetic field energy density that is calculated in the upstream fluid frame. This assumption is not universally valid, as reconnection at high $\sigma$ leads to significant changes to fluid velocity inside the current sheet where reconnection occurs. For example, it has been shown that the velocity component along the upstream field does not lead to a change in the reconnection rate as long as it is significantly lower than the Alfven speed \citep{Hakobyan2023psr}. In the case considered here, however, the Lorentz factors in the upstream are mild, $\Gamma\lesssim {\rm few}$, so this correction is not significant. We defer issues related to the dependence of reconnection rate on upstream fluid velocity to a future dedicated study.}. The superphotons originating close to the black hole have larger weight because of the radial dependence of $b^2$ (see Figure \ref{fig:bsq_profile}). Reconnection and, thus, photon emission, happens along the jet base. Unlike in two-dimensional simulations \citep{2020ApJ...900..100R}, the current sheet is no longer located dominantly in the equatorial plane for flux eruptions in three dimensions. For strongly beamed models that are particularly sensitive to the geometry of the magnetic field in the emission region, this fact highlights the importance of capturing fully 3D dynamics of the eruption for VHE lightcurve modeling.  

\begin{figure}
    \centering
    \includegraphics[width=0.49\textwidth]{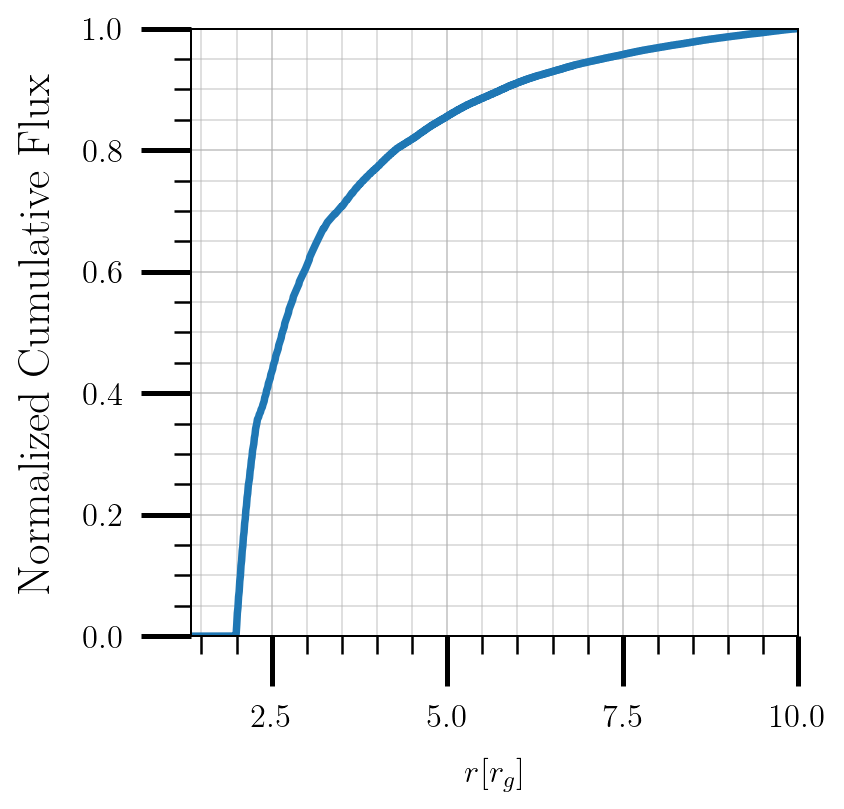}
    \caption{The normalized cumulative distribution of the total luminosity from the current sheet as a function of the radial coordinate measured in $r_{\rm{g}}$. The radial dependence of luminosity is identical for all of the beaming models. About $80\%$ of the total flux originates within $5 r_{\rm{g}}$ from the black hole. This is because of the strong radial dependence of the VHE emissivity ($\propto b^2 \sim r^{-3.5}$).}
    \label{fig:luminosity_radial_dependence}
\end{figure}

We plot the cumulative VHE flux as a function of distance from the black hole in Figure~\ref{fig:luminosity_radial_dependence}. The cumulative flux at position $\mathbf{r}$ is obtained by computing the normalized sum of weights of all of the superphotons at $10^4$ $r_{\rm g}$, which have an initial Kerr-Schild coordinate of $r \leq \mathbf{r}$. We find that there are almost no superphotons in the inner $\approx 2$ $r_{\rm{g}}$ that contribute to the flux that makes it to the outer boundary. About $80\%$ of the flux comes from $r \lesssim 5$ $r_{\rm{g}}$ of the black hole. This fact stems from the larger magnetic field energy density available to power the IC emission. The cumulative VHE flux as a function of radius is very similar for all of the beaming models (i.e., cumulative fluxes constructed for different values of $\Delta \chi$), and here we only show the isotropic case. 

\begin{figure*}
    \centering
    \includegraphics[scale=0.6]{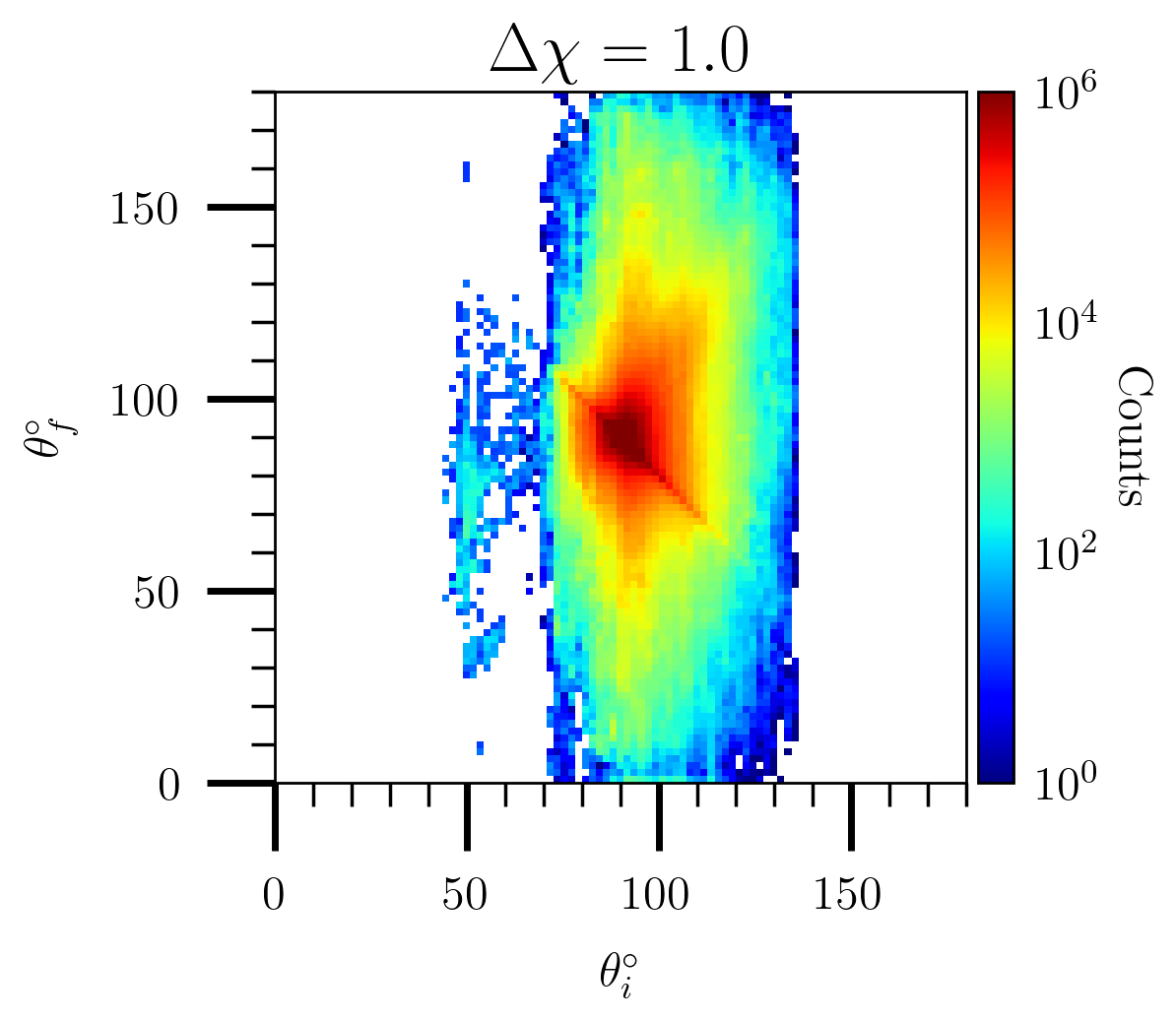}
    \includegraphics[scale=0.6]{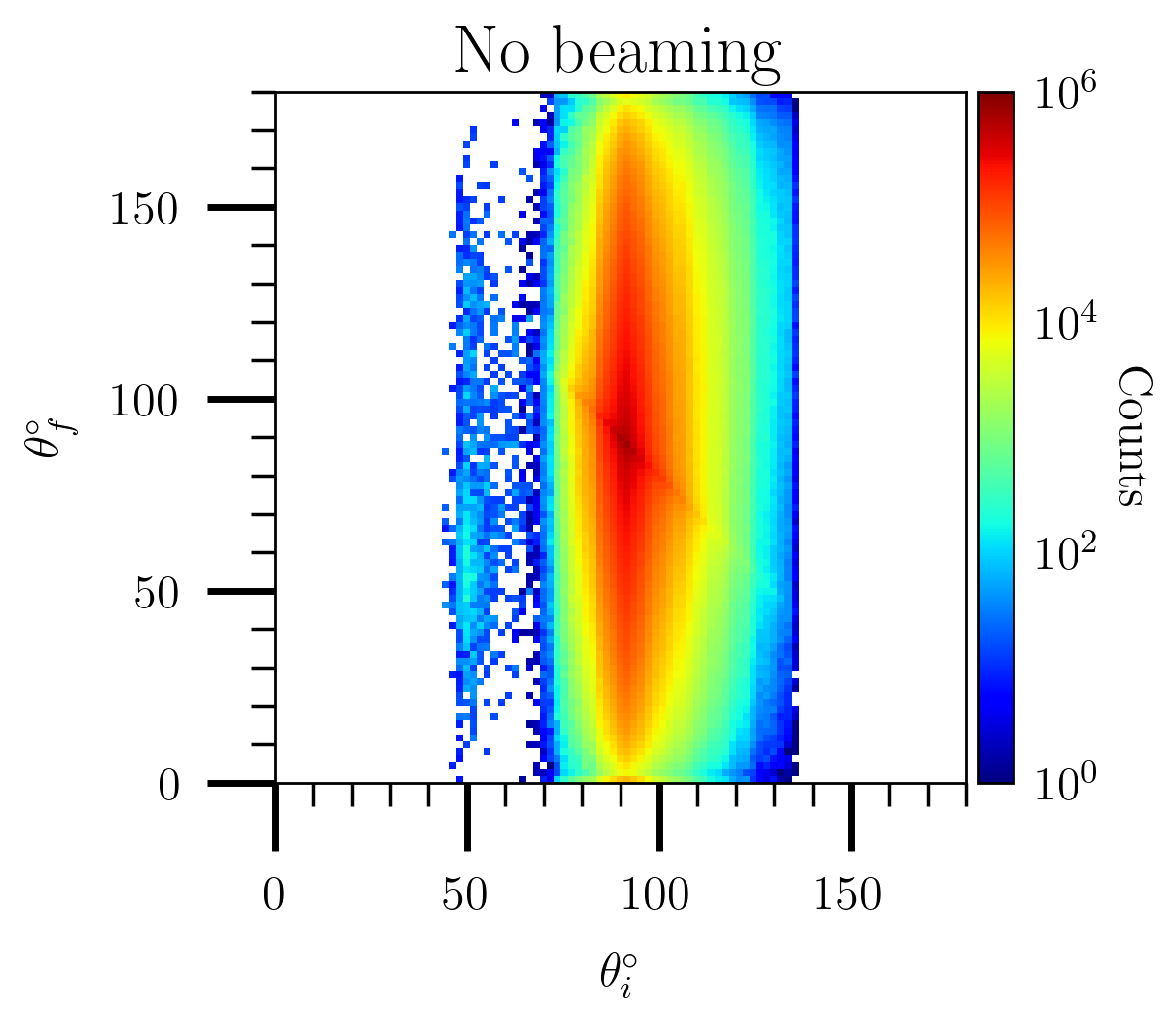}
    \caption{$2-$D histograms for the initial and final $\theta$ coordinates of superphotons in the lightcurves computed for strong and no beaming cases. Most of the initial emission is concentrated in a small range of angles, namely $80^{\circ} < \theta_i < 120^{\circ}$ for both cases. In the strong beaming case $(\Delta \chi = 1^{\circ})$, the superphotons stay close to the equatorial plane even at large distances. On the contrary, superphotons propagate more uniformly when there is no strong beaming assumed, which is as expected. This is because in the vicinity of the black hole, the upstream magnetic field points in the $\hat{e}_r-\hat{e}_{\phi}$ direction, where $\hat{e}_r $ and $\hat{e}_{\phi}$ are unit vectors along $r$ and $\phi$ directions respectively. Therefore, the emission is beamed along the same direction. See the text in \ref{Results: emitting regions} for the tight correlation between $\theta_i$ and $\theta_f$. }
    \label{fig:theta_histogram}
\end{figure*}

Figure \ref{fig:theta_histogram} shows the 2-dimensional histograms of the initial and final (computed at $10^4 r_{\rm g}$) $\theta$ coordinates of the superphotons, for the very strong, $\Delta \chi = 1^{\circ}$, and no beaming models, respectively. Their initial and final values are given by $\theta_i$ and $\theta_f$, respectively. Most of the calculated flux originates from the vicinity of the equatorial plane, $\theta_i \in \{80^{\circ}, 120^{\circ} \}$. In the case of very strong beaming, the outgoing photons stay in the equatorial plane, and therefore the spread in $\theta_f$ is small. On the other hand, in the model with uniform beaming, the flux is more uniformly distributed over all latitudes. There is also a tight correlation between the initial and final arrival latitudes of superphotons such that $\theta_f = 180^{\circ} - \theta_i$. This is clearly evident in the strong beaming model, where most of the superphotons are initialized close to the equatorial plane and along the magnetic field lines. Therefore, they move in the $r$ direction towards or away from the black hole. This trend is also visible in the no-beaming model, although to a lesser extent. This correlation can be understood as follows. In addition to the energy, norm of the wavevector and angular momentum, the geodesics also have an additional conserved quantity called the Carter's constant. In Boyer-Lindquist (BL) coordinates ($t_{BL}, r_{BL}, \theta_{BL}, \phi_{BL}$), this quantity can be written as $\mathcal{C} = k_{\theta_{BL}}^2 + \cos^2\theta_{BL} (k_{t_{BL}}^2 a^2 - k_{\phi_{BL}}^2 / \sin^2 \theta_{BL})$, where $\mathcal{C}$ is the Carter's constant and $|k_{t_{BL}}|$ and $k_{\phi_{BL}}$ are the conserved energy and angular momentum respectively. The initial wavevectors of the superphotons have a negligible $k^{\theta_{BL}} = k^{\theta}$ component because they follow the upstream magnetic field distribution, which is radially dominated (note that $k^{\theta_{BL}}$ is equivalent in both the coordinate systems because $\theta_{BL} = \theta$); see Appendix \ref{Appendix: selection of the upstream magnetic field}. Then $k_{\theta} \propto k^{\theta}$ vanishes in BL coordinates, and therefore the superphotons initially are at a turning point in $\theta$. The outer boundary is also a turning point in $\theta$ since the superphotons propagate radially outwards at infinity and $k^{\theta} \rightarrow 0$. The second turning point is located at $\theta_f = \pi - \theta_i$, which can be verified by substitution into the equation for the Carter's constant.

\subsection{Flare Lightcurves}\label{Results: flare lightcurve}
\begin{figure*}
    \centering
    \includegraphics[width=0.7\textwidth]{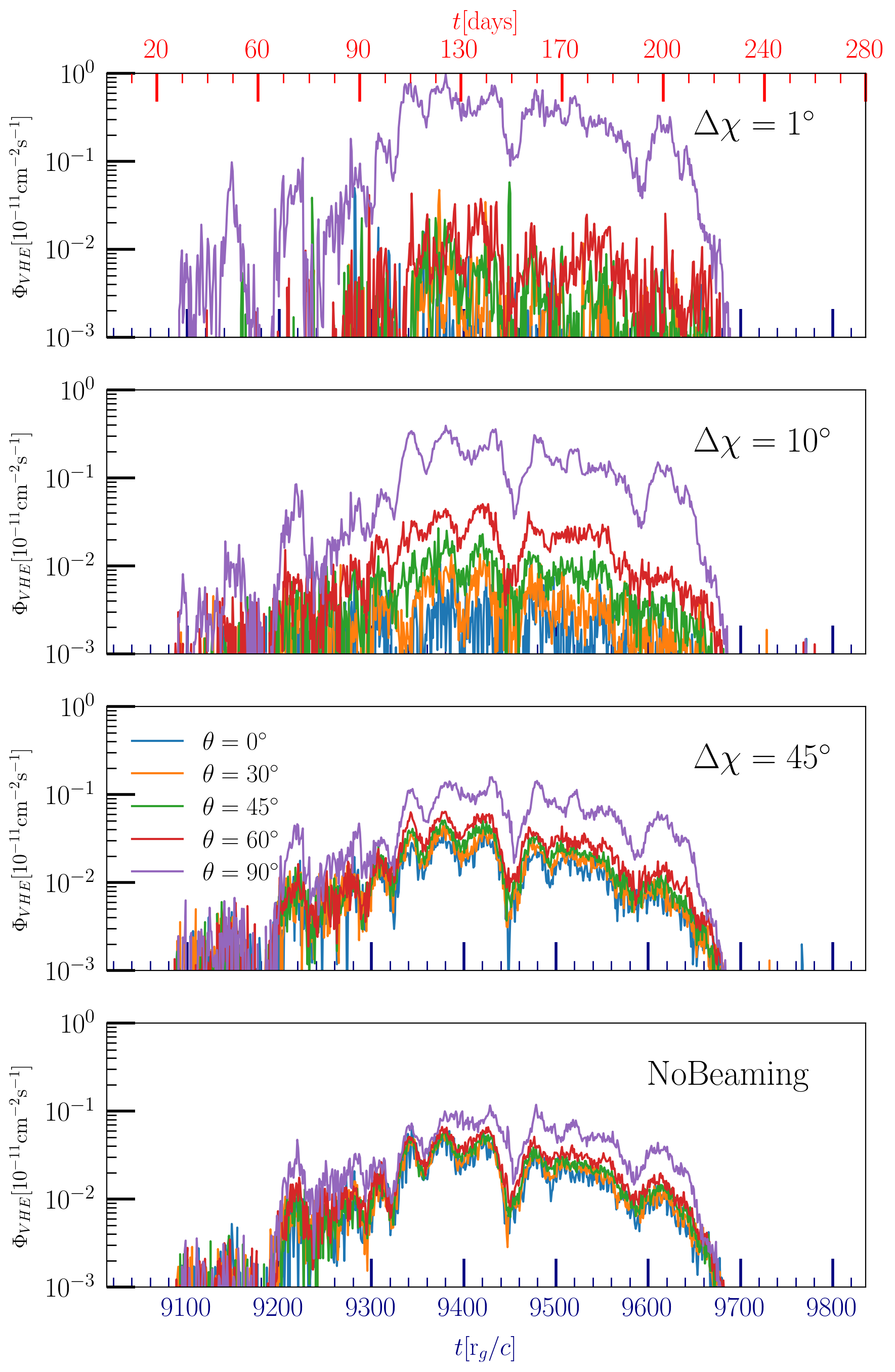}
    \caption{Lightcurves for different photon beaming models obtained from binning the Kerr-Schild coordinate times of the superphotons at $10^{4} r_{\rm{g}}$. $\Delta \chi$ is a measure of the beaming strength, with a smaller $\Delta \chi$ corresponding to superphotons aligned to the upstream magnetic field. The colors in each plot represent solid angles at different inclinations over which the flux is received. In the extreme beaming case $\Delta \chi = 1^{\circ}$, most of the received emission arrives in the equatorial plane of the simulation. There is approximately a factor of $100$ suppression in the luminosity at higher latitudes. This fraction changes once $\Delta \chi$ gets larger, that is, there is less beaming. In the other extreme case where there is no beaming, there is only a factor of a few difference in received flux near the equator and at higher latitudes. The superphoton count is first normalized by re-scaling all of the lightcurves by the median flux in the no-beaming model. This flux is then equated to the corresponding flux arising from a flare with luminosity of $\sim  10^{40}$ erg s$^{-1}$ at a distance of $16$ Mpc. The photon count is obtained by dividing the flux by a characteristic photon energy of $2$ TeV to compare with VHE observations in \cite{2012ApJ...746..151A}.}
    \label{fig:lightcurves}
\end{figure*}

We plot the lightcurves for models with different beaming in Figure \ref{fig:lightcurves}. The superphoton wavevectors are integrated out to $10^4 r_{\mathrm{g}}$ and their coordinates are stored. The lightcurves are constructed by creating a histogram of the superphotons' time coordinates once they reach the outer boundary, weighted by the photon weight, $w_i$. Each color represents the total flux at different inclination ranges with respect to the spin axis of the black hole in the simulation. The inclination ranges are chosen such that an equal amount of solid angle is contained in every curve.

The superphoton flux further requires to be normalized to compare with the VHE observations. To do so, we equate the median flux over all solid angles to a fiducial luminosity of $10^{40}$ ergs s$^{-1}$; see Section \ref{VHE-IC:particle_energization}. Then, the flux at time $t$ going into a solid angle $\Delta \Omega = 2 \pi  \int_{\theta_i}^{\theta_f} \sin \theta d\theta$ is given by:
\begin{align}
    F(t, \theta=\theta_i) = 10^{40} \cdot \frac{\int_{\theta_i}^{\theta_f} \Delta N(t,\theta_i) \sin \theta d\theta}{\overline{\Delta N} \Delta \Omega d_{\rm M87}^2} \\
    \text{ ergs s}^{-1} \text{cm}^{-2} \nonumber,
\end{align}
where $\Delta N(t, \theta)$ are the number of (weighted) superphotons received between times $t$ and $t + t_g$, and angles $\theta_i$ and $\theta_f$, and $L_0=10^{40}{\rm{erg/s}}$ is the fiducial IC luminosity (Section 2.2). Here, $\theta_f$ is computed so that an equal amount of solid-angle, taken to be $\Delta \Omega \equiv 2 \pi \times (\pi / 180)$, is contained for all lightcurves, so that $\theta_f = \arccos \Bigl(\cos \theta_i -  \frac{\Delta \Omega}{2\pi} \Bigr)$. In all of the constructed lightcurves, $\theta_f$ differs from $\theta_i$ by less than $1^{\circ}$. The normalization constant is given by $\overline{\Delta N}$, which is the $\text{median value of } \int_{0}^{\pi} \Delta N(t,\theta_i) \sin \theta d\theta$ in time.
\newline
Since our ray tracing does not keep track of the superphoton energies, we need to relate the superphoton count to the VHE flux \footnote{Note that we construct the lightcurves using an optical depth of $\tau = 0$, i.e., there is no attenuation in the signal during propagation. Estimates for the VHE photon optical depth for M87* vary in the literature. $\tau \lesssim 1$ is expected for energies less than a TeV \citep{Broderick_S_2015ApJ...809...97B,curvature_2024A&A...685A..96H}}. Therefore, we adopt a photon power-law index of 2.2 starting from 350 GeV \citep{power_law_index_2012ApJ...746..141A}. The average superphoton energy for such a distribution is about $ 2 \text{ TeV} \approx 3.2$ ergs. The flux is then converted into counts s$^{-1}$ cm$^{-2}$ using this value. The fiducial luminosity, $L_0$, further scales with the assumed strength of the magnetic field near the black hole, properties of the soft photon field, and distribution of accelerated particles (see Section 2.2). 

The maximum flux from the beamed lightcurves gets as large as $\sim 10^{-11}$ photons cm$^{-2}$ s$^{-1}$, which is suppressed by a factor of $10$ for the no-beaming lightcurves. The VHE flux is largest at the equator ($\theta_i = 90^{\circ}$) for all of the lightcurves, which results from two different beaming effects at play.  Most of the VHE flux comes from the inner part of the current sheet, which lies roughly in the equatorial plane of the simulation. In this region, the upstream magnetic field lines, and therefore the wavevectors of superphotons in the strong beaming model, are oriented in the radial direction. Because of this fact, most of the flux ends up along the equatorial plane. A high-altitude observer will only see the emission if the upstream magnetic field points out of the equatorial plane. Consequently, the emission drops by $2$ orders of magnitude at higher latitudes in the lightcurves with $\Delta \chi = 1^{\circ}$ and $10^{\circ}$. The lightcurves at these high latitudes show also a faster timescale variation of less than a day, although the overall flux is highly suppressed.

\begin{figure*}
    \centering
    \includegraphics[width=0.99\textwidth]{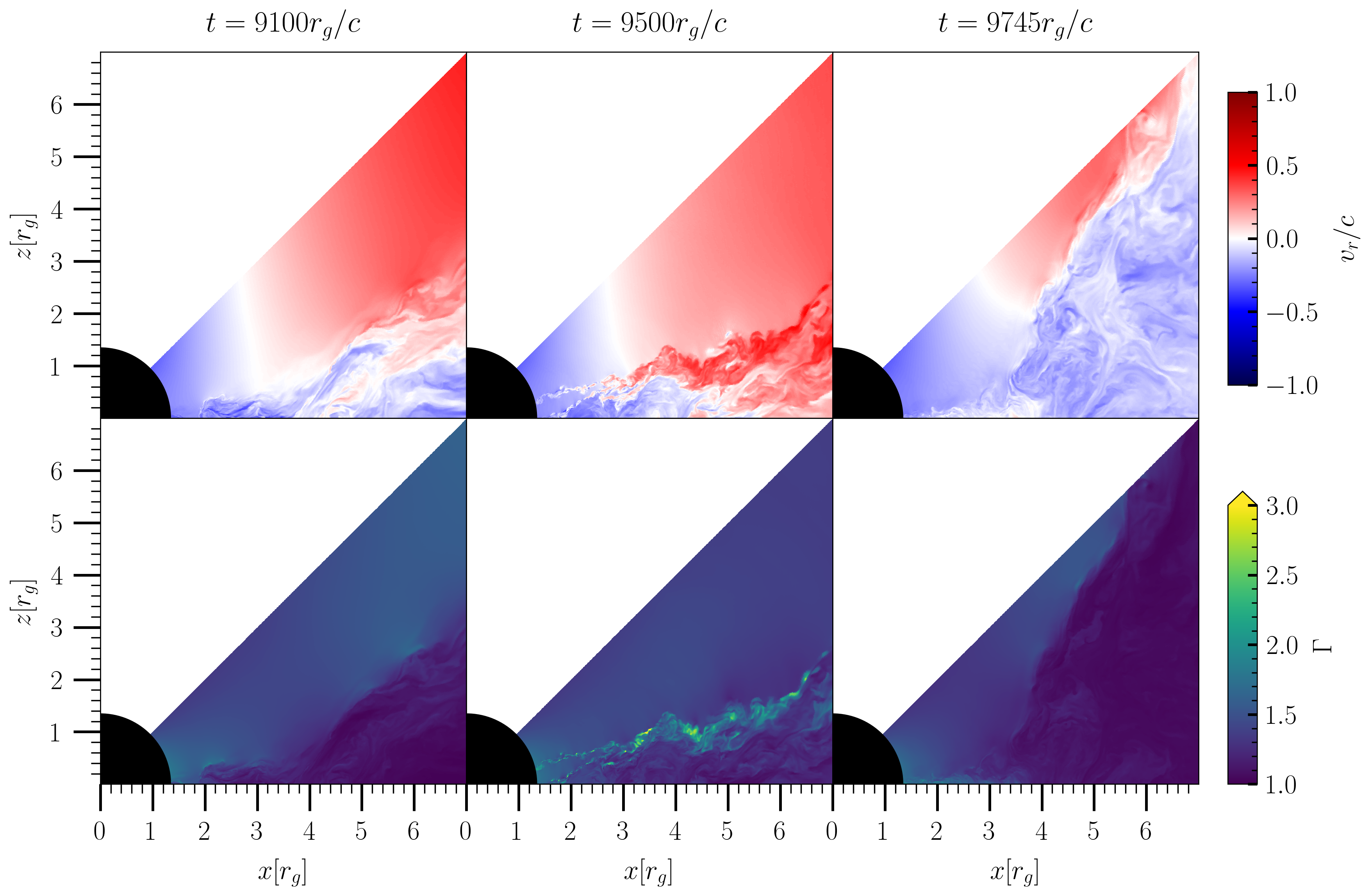}
    \caption{The radial velocity and the Lorentz factor of plasma near the current sheet. Relativistic beaming from the bulk motion of the plasma is another effect that can beam the IC flux in the equatorial plane. The Lorentz factor of plasma near the equatorial plane is $\Gamma \lesssim 2$, which moves radially outwards during the flux eruption. As a result, there is an equatorial beaming in the lightcurve that is $\propto \Gamma^2 \lesssim 4$, even if no beaming is assumed.}
    \label{fig:lfac_radial_vel}
\end{figure*}

Additionally, there is an increased flux along the equatorial plane, even in the case of no beaming. This is a result of the mildly-relativistic bulk plasma motion in the radial direction close to the black hole. Figure \ref{fig:lfac_radial_vel} shows the radial 3-velocity and the Lorentz factor of the plasma during the magnetic flux eruption. The increased equatorial flux by a factor of a few can be explained by the bulk motion of the plasma. The beaming in this case is proportional to $\Gamma_{\mathrm{upstream}}^2 \lesssim 4$ and not as strong of an effect as the assumed beaming.

\begin{figure*}[h]
    \centering
    \includegraphics[width=0.7\textwidth]{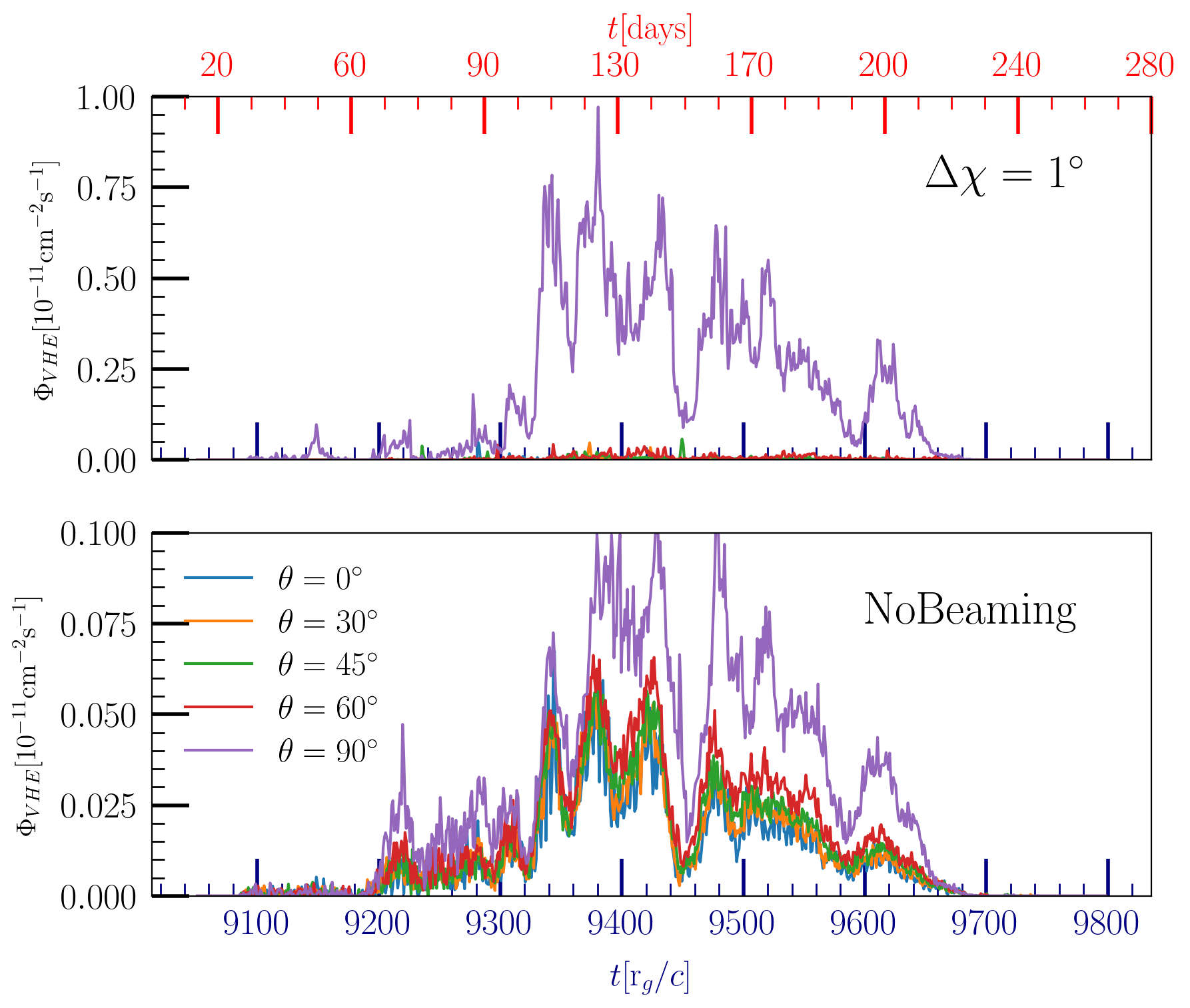}
    \caption{Same lightcurves as in Figure \ref{fig:lightcurves} but plotted on a linear scale on the $y$-axis to better compare with the observations. The superphoton counts are converted to a photon number flux using the same procedure as in Figure \ref{fig:lightcurves}. Although the entire flare lightcurve lasts for over $100$ days, shorter timescale features are still visible, resulting from the changing $4-$volume of the current sheet.}
    \label{fig:lightcurves_linear}
\end{figure*}

In order to compare the lightcurves with the observations, we plot the strong beaming case, $\Delta \chi=1^{\circ}$, and isotropic models, on a linear scale in Figure \ref{fig:lightcurves_linear}. Assuming a  threshold of $10^{-11} \text{ cm}^{-2} \text{ s}^{1}$ (e.g., the threshold flux for the very strong VHE flares in \cite{2012ApJ...746..151A}), the flux from the lightcurve with strong beaming would not appear as a very energetic VHE flare at higher latitudes. However, when there is no beaming, the flux at high latitudes is only suppressed by a moderate factor of $\sim 2-3$. 

\begin{figure*}
    \centering
    \includegraphics[width=0.99\textwidth]{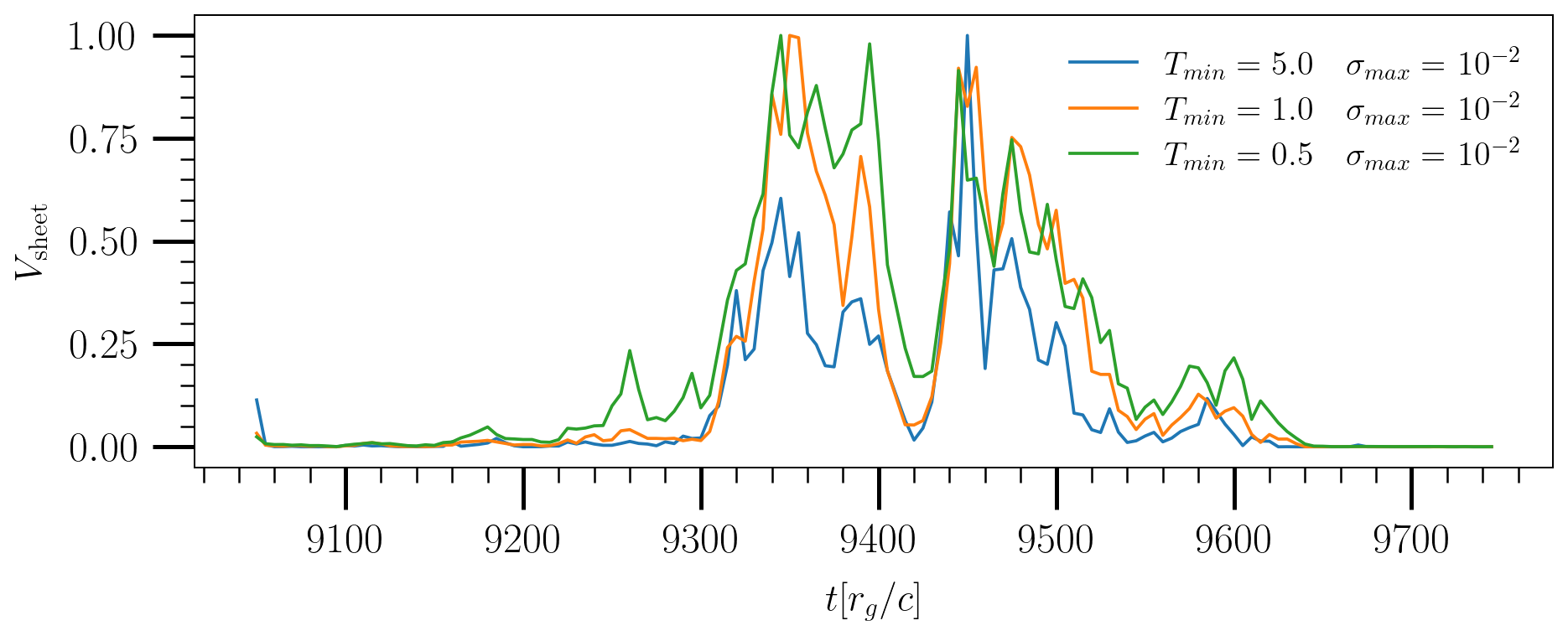}
    \includegraphics[width=0.99\textwidth]{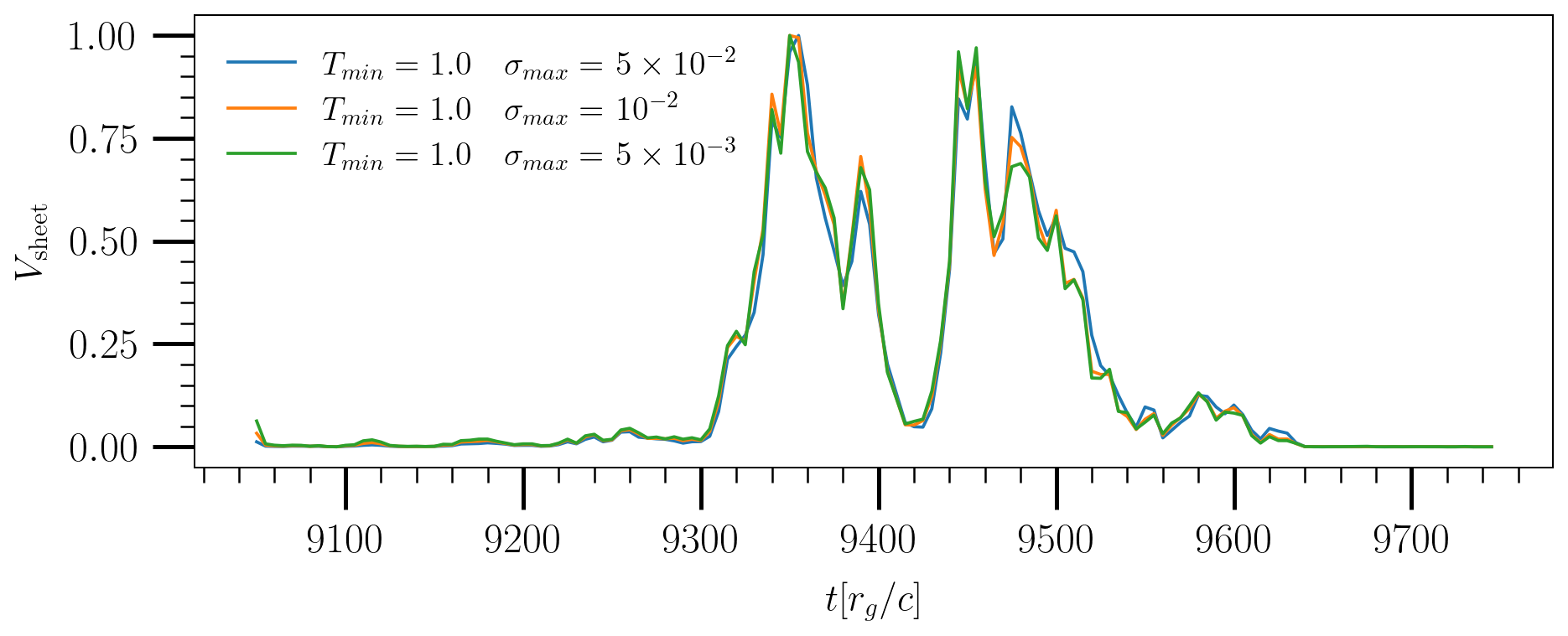}
    \caption{The re-scaled volumes of the current sheets identified using different $(T,\sigma_c)$ thresholds as a function of the Kerr-Schild coordinate time. We separately vary the $T$ and the $\sigma_c$ thresholds to identify the current sheets and calculate the $4$-volumes. The respective volumes are computed by summing up $dV_i = \sqrt{-g_i}d^4x$ of the current sheet, where $dV_i$ is the 4-dimensional volume of the current sheet cell $i$. They are normalized by dividing by the maximum volume attained during the flux eruption. Results from different $T$ thresholds are shown in the left panel and for different $\sigma_c$ thresholds in the right panel. Orange curves represent the default $T=1$ and $\sigma_c = 10^{-2}$ thresholds used in the analysis in this paper. There are some differences in the relative change in $4-$volumes while varying the temperature, but these differences occur on similar timescales. Varying $\sigma_c$ produces very similar results. The $4$-volumes all show shorter-timescale features that are related to the global, non-linear dynamics of the system. These variations are imprinted on the VHE lighcurves in Figure \ref{fig:lightcurves_linear}.}
    \label{fig:current_sheet_volume}
\end{figure*}

While the entire flux eruption, indicating the period in which a flare could be powered in the simulation, lasts for $\sim 400 r_{\mathrm{g}}/c$ ($120$ days), there is a much faster variability present in all the lightcurves, on timescales of $\sim 20$ $r_{\mathrm{g}}/c$ ($15$ days). This variability closely resembles the changing $4-$volume of the current sheet over time, which traces the total amount of the dissipated power. It is important to check how the volume of the identified sheet changes when it is identified using different $(T,\sigma_c)$ thresholds. We plot the normalized $4-$volumes of the current sheet identified using different $(T,\sigma_c)$ thresholds in Figure  \ref{fig:current_sheet_volume} as a function of time \footnote{Note that we use the $4-$volume instead of the $3-$volume to also account for the change in flux from gravitational time dilation.}. It is computed as $V_{\rm{sheet}} = \sum_{i} \sqrt{-g_i}$, where $V_{\rm{sheet}}$ is the $4-$dimensional volume, and the summation is done over all the cells in the sheet (note that the measure $\Delta ^4 \mathrm{x^{\mu}}$ is constant for each cell and is not included in the 4-volume calculation). The volume is normalized by its maximum value during the flux eruption. We conclude that our results are not very sensitive to the choice of $T$ and $\sigma_c$: for any threshold, the volume changes on a $\sim 20  r_{\mathrm{g}}/c$ timescale, which gets imprinted in the lightcurve. 

From the current sheet identification, we find that there is always a small amount of persistent magnetic reconnection near the event horizon before the onset of the flux eruption. We take this as a measure of the quiescent flux in our lightcurves. Taking the sudden increase in the flux at $t = 9350 r_{\mathrm{g}}/c$ to be the start of the flare (which is consistent with the rapid increase in the $4-$volume of the current sheet at $t=9300 r_{\text{g}}/c$), Figure \ref{fig:lightcurves_linear} shows a factor of $5-10$ increase in flux during the flare in the model without strong beaming.

\vspace{2cm}
\section{Discussion} \label{Discussion}
\subsection{Key Results}\label{discussion:key_results}
We constructed VHE lightcurves of M87* by post-processing a GRMHD simulation of an accretion disk in a magnetically arrested state. Based on prescriptions from self-consistent kinetic simulations of a strongly cooled reconnecting pair plasma, we initialized and ray traced photons from a reconnecting current sheet. We showed that the lightcurves depend on both the global dynamics and the microphysical prescriptions for the particle acceleration. 

The lightcurves generically show a long-term variability during which the flux rises by over a factor of $10$. This timescale is as long as the duration of the magnetic flux eruption in the simulation, which is about $400$ $r_{\mathrm{g}}/c$, or about $130$ days for M87*. The variability is modulated by faster `sub-flares’ that last for $\sim 40$ $r_{\mathrm{g}}/c$, or a little over $10$ days. The fractional flux changes by a factor of $2-5$ in the sub-flares, which occur about $10$ times during the flux eruption. The duration of the sub-flares correlates with the timescale of the changing volume of the current sheet, which is set by the dynamics of the current sheet. 

Beaming and variations in the current sheet extent can boost the observed flux in the lighcurves of strongly beamed models, $\Delta \chi = 1^{\circ}$, compared to the no-beaming lightcurves. This can account for the observed flux of $\sim 10^{-11}$ photons cm$^{-2}$ s$^{-1}$ in the most energetic flares, even if the median flare luminosity is of the order $\sim 10^{40}$ erg s$^{-1}$. The necessity of beaming in the reconnection model motivates conducting detailed investigations of the relative orientation of the current sheet and the black hole magnetic spin-axis in different accretion models.  

Note that the GRMHD simulation shows multiple smaller scale flux eruptions, resulting in short-lived and shorter current sheets, which we have not ray traced here. We focus on the largest flux eruptions, producing the flares with the longest time-duration.


\subsection{Comparision with Observations}\label{discussion:comparision_with_observations}

For all of our models, most of the flux in the lightcurves originates from the inner $5 r_{\rm g}$, where the current sheet dominantly lies in the equatorial plane. Consequently, the flux in the lightcurves of strongly beamed models is highly concentrated in the equatorial plane because the upstream magnetic fields are ordered in the toroidal direction. For a nearly jet-aligned observing angle expected in M87*, we conclude that our simulations are not providing enough variability in the current sheet orientation to explain the observed flares in the strong beaming model, which corresponds to the very strong cooling. However, the sub-flares observed in models with moderate degree of beaming can be potentially reconciled with the data. The VHE flares from M87* show flux variations that are as short as $2$ days. The sub-flares can get close to these short timescales and have $\Delta F / F \sim 2-5$, similar to the fractional flux variations of the observed flares. 

If we consider the observed M87* VHE flares to represent the sub-flares within a longer flaring episode, we predict that the sub-millimeter and VHE lightcurves should vary on different timescales. The ejection of the accretion disk during the flux eruption can produce a characteristic sub-millimeter dimming of the $230$ GHz flux on the same timescale as the flux eruption \citep{Jia_2023arXiv230109014J}. While the baseline VHE flux should increase during this longer timescale, it can still vary much quicker because of the shorter duration of the sub-flares, which is tracing the evolution of the current sheet during the flare. Long-term multi-wavelength monitoring of M87* can provide more insight into the relation of the sub-millimeter and VHE emission properties during flux eruptions.

\subsection{Caveats}\label{discussion:caveats}
While our methods of constructing VHE lightcurves are motivated by self-consistent kinetic simulations, they contain several constraining assumptions that can be addressed in future work. 

The plasma around M87* is collisionless, and first-principles modeling of reconnection dynamics requires fully kinetic simulations. In particular, axisymmetric GRPIC simulations showed that timescale of magnetic flux eruptions can be shorter as a result of larger rate of magnetic reconnection  in kinetic simulations \cite{Alisa_2023PhRvL.130k5201G}. This may well apply to the timescales for the variations in the current sheet volume in $3$-dimensional kinetic simulations, which we identified as one of the important sources of the VHE lightcurve variability during the eruption. This is particularly relevant because the shortest timescales in our ray traced lightcurves are still a factor of few longer compared to those observed in VHE flares. 

A related source of uncertainty is connected with the geometry of flux eruptions. In particular, it remains to be seen whether more tilted current sheet geometries can be obtained in more realistic accretion setups that do not start with much-simplified rotating plasma tori (e.g., for tilted disks \citep{whitequataert2019ApJ...878...51W,chatterjeetilted2023arXiv231100432C} or in accretion flows with less ordered angular momentum \citep{ressler2023MNRAS.521.4277R,alisa2024arXiv240911486G}). If more inclined current sheets can be realized, models with strong beaming could produce enough variability while still producing enough flux for nearly face-on viewing angles. 

To construct the lightcurves, we assumed the same level of the beaming of IC radiation throughout the current sheet, using order of magnitude estimates for $\gamma_{\rm{rad}}$ and $\sigma_c$. Plasma magnetization, $\sigma_c$, is particularly uncertain because it is supposed to be set by the self-consistent regulation of pair density produced by high-energy synchrotron photons emitted by accelerated particles, and the global infow-outflow effects. Local and global radiative kinetic simulations that can capture these effects are important to produce more reliable estimates of $\sigma_c$. Because of these uncertainties, we constructed lightcurves for both the strong cooling, $\sigma_c>\gamma_{\rm rad}$ (strong beaming), and weak cooling, $\sigma_c \sim \gamma_{\rm rad}$ (no beaming),  regimes. For similar reasons, we do not attempt to construct energy-dependent lightcurves with different levels of beaming.

VHE observations of M87* constrain its optical depth $\tau \leq 0.2$ for $E \leq 10$ TeV \citep{curvature_2024A&A...685A..96H}. Therefore, our ray tracing does not include the effects of $\gamma$-ray attenuation, i.e., we assume $\tau = 0$). This assumption remains to be verified theoretically. In particular, the nominal $\gamma-\gamma$ pair-production optical depth can be estimated to be $\sim 1$ for $1$ TeV photons \citep{Hayk_2023ApJ...943L..29H}, assuming a quiescent soft-photon radiation field observed from M87* \citep{Broderick_S_2015ApJ...809...97B}. This indicates that pair-production attenuation can be non-negligible. However, the sub-millimeter flux from the inner accretion zone can be suppressed during the magnetic flux eruption \citep{Jia_2023arXiv230109014J}, so that more detailed calculations are needed.

\section{Conclusions} \label{sec:Conclusions}
Current sheet formation is ubiquitous in many astrophysical systems containing black holes and neutron stars. These are the sites of particle energization and production of the high-energy emission. However, first-principles modeling of the high-energy lightcurves from these systems can be challenging because of the large separation of scales between the Larmor radii of the accelerated particles and the system size. In this work, we describe a general method, which is applicable whenever the cooling length of high-energy emitting particles is short, such that they do not need to be traced through the global scales. We use the results from local kinetic simulations that describe the beaming of accelerated particles, combined with GRMHD simulations that capture the global dynamics, and in particular the geometry of the current sheet and upstream magnetic field. This procedure can be similarly applied to other astrophysical systems with different values of plasma magnetization, $\sigma_c$, and radiative cooling, $\gamma_{\rm rad}$, parameters; different levels of beaming can be modeled based on the ratio of the two parameters.

The production of VHE emission from low-luminosity AGN is not well understood. We explore the combined role of kinetic (beaming of accelerated particles) and global (current sheet dynamics), including general-relativistic (photon orbits), effects on the VHE lightcurves by ray tracing GRMHD simulations of a rapidly spinning black hole undergoing large-scale magnetic flux eruptions. We find: (a) most of the emission originates from very close $(r \lesssim 5 r_{\rm{g}})$ to the black hole; (b) the duration of the observed flare  is roughly equivalent to the timespan of the magnetic flux eruption; (c) there are multiple sub-flares within the main flare that occur as a result of the changing volume of the current sheet, and this result is insensitive to the numerical procedure of the current sheet identification; and (d) there can be over a factor of $50$ suppression of outgoing flux away from the equatorial plane in all of the lightcurves because of the IC beaming. Future VHE flare observations in multiple energy bands could better inform us about the parameters of the flaring plasma around the black hole M87*.
\section*{Acknowledgements} \label{Acknowledgements}

The authors thank Amir Levinson and Chris White for fruitful discussions. S.S., B.R. and A.P. are supported by a grant from the Simons Foundation (MP-SCMPS-00001470). J.D. is supported by NASA through the NASA Hubble Fellowship grant HST-HF2-51552.001A, awarded by the Space Telescope Science Institute, which is operated by the Association of Universities for Research in Astronomy, Incorporated, under NASA contract NAS5-26555. B.R. is supported by the Natural Sciences \& Engineering Research Council of Canada (NSERC). A.P. additionally acknowledges support by NASA grant 80NSSC22K1054, Alfred P.~Sloan Research Fellowship and a Packard Foundation Fellowship in Science and Engineering. This research was facilitated by the Multimessenger Plasma Physics Center (MPPC), NSF grant PHY-2206607. This research was enabled by INCITE program award PHY129, using resources from the Oak Ridge Leadership Computing Facility, Summit, which is a US Department of Energy office of Science User Facility supported under contract DE-AC05- 00OR22725 and, as well as Calcul Quebec (http://www.calculquebec.ca), Compute Canada (http://www.computecanada.ca) and Compute Ontario and the Digital Research Alliance of Canada (alliancecan.ca). The computational resources and services used in this work were partially provided by facilities supported by the Scientific Computing Core at the Flatiron Institute, a division of the Simons Foundation; and by the VSC (Flemish Supercomputer Center), funded by the Research Foundation Flanders (FWO) and the Flemish Government – department EWI. This research is part of the Frontera \citep{Frontera} computing project at the Texas Advanced Computing Center (LRAC-AST21006).

\bibliography{bibliography}{}
\bibliographystyle{aasjournal}
\appendix

\section{Thresholds for Current Sheet Identification} \label{Appendix: thresholds for current sheet identification}

\begin{figure}
    \centering
\includegraphics[width=0.99\textwidth]{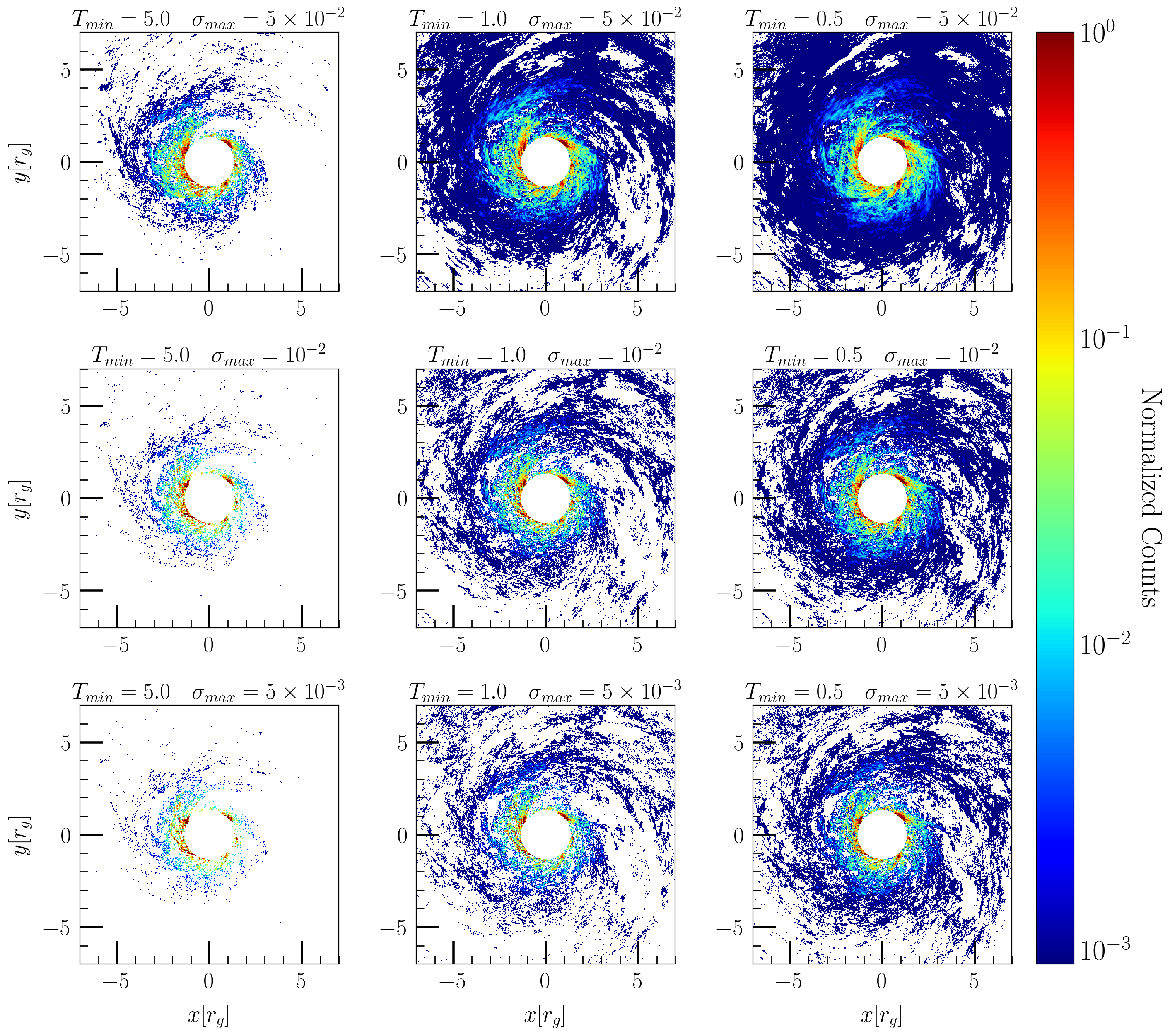}
\caption{The location of the current sheet, identified using various $(T,\sigma_c)$ thresholds. We plot the binned location of the current sheet zones after projecting them on the equatorial plane in cartesian Kerr-Schild coordinates, weighted by $r^{-3.5} \propto b^2$. The weighting is proportional to the fluid frame magnetic energy density that is proportional to the IC emissivity. The location of the current sheet close to the black hole ($r \lesssim 5 r_{\mathrm{g}}$) is similar for all of the thresholds, where most of the flux originates in our lightcurves.}
\label{fig:current_sheet_thresholds}
\end{figure}

It is important to check how sensitive the identified current sheet, and therefore, the lightcurves are to the $T$ and $\sigma_c$ thresholds. To that end, we vary both thresholds, identify the corresponding current sheet, and plot the results in Figure \ref{fig:current_sheet_thresholds}. The minimum (dimensionless) temperature thresholds are taken from $\{0.5, 1, 5\}$ and the maximum $\sigma_c$ thresholds from $\{5\times 10^{-3}, 10^{-2}, 5\times 10^{-2}\}$. The $T=1$ and $\sigma_c = 10^{-2}$ thresholds are used in the rest of the paper. The process is repeated for all of the simulation outputs and the coordinates of the current sheet cells are combined and binned into the $x-y$ coordinates. $x,y,z$ represent the cartesian Kerr-Schild coordinates that can be obtained from the standard transformation of spherical into cartesian coordinates.

The binned data are weighted by $r^{-3.5}$ to account for the emissivity that is proportional to the strength of the upstream magnetic field in the fluid frame (see \ref{subsec: photon initialization}). The outer extent of the current sheet reduces for the $T=5$ threshold. However, in all of the cases, the current sheet structure in the inner $5 r_{\mathrm{g}}$ is very similar, where most of the emission originates in the lightcurves. We therefore conclude that $(T,\sigma_c) = (1, 10^{-2})$ are reasonable choices of thresholds to use for the current sheet identification.

\section{Selection of the upstream magnetic field} \label{Appendix: selection of the upstream magnetic field}
\begin{figure}
    \centering
    \includegraphics[scale=0.6]{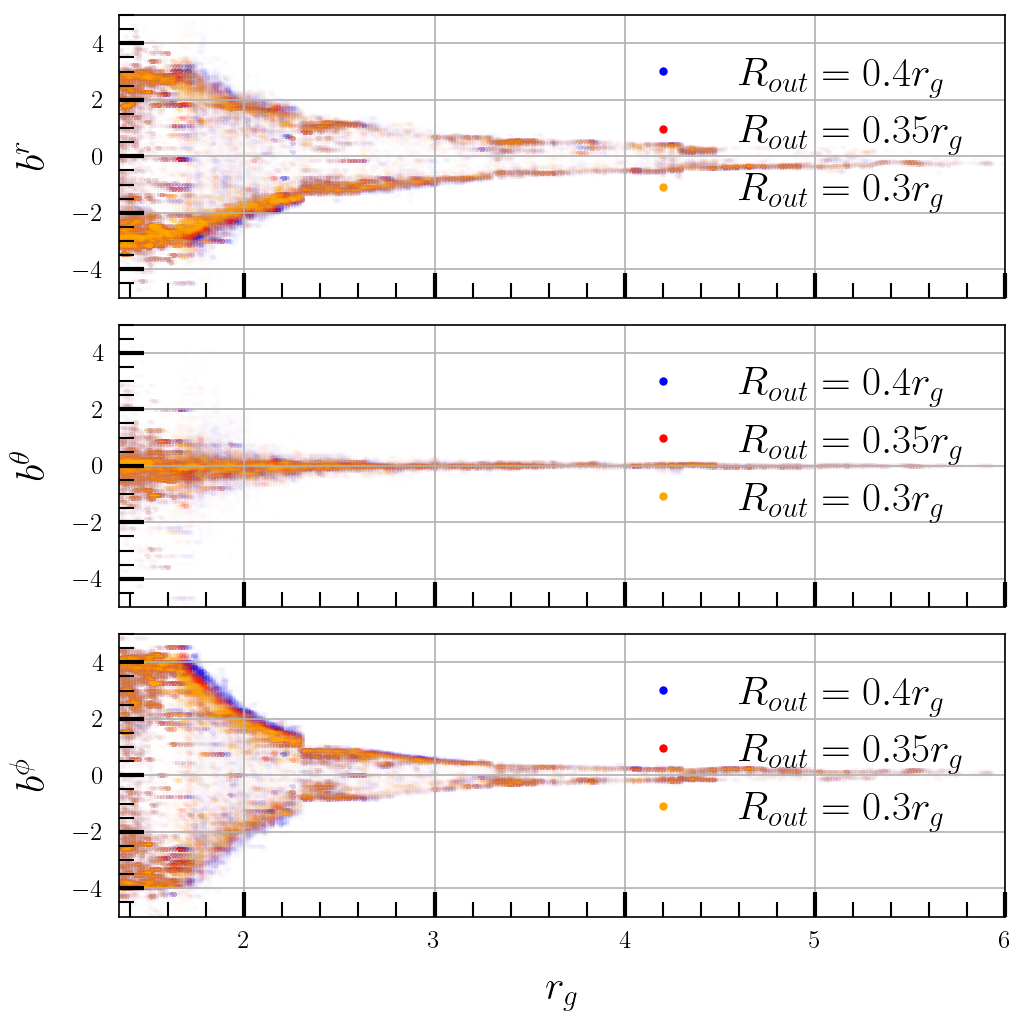}
    \caption{Figure \ref{fig:bcorner_plot} shows $2$D histograms of the upstream magnetic field components with respect to the radial coordinate $r$. The panels from top to bottom show $b^r$, $b^{\theta}$ and $b^{\phi}$ of the upstream field, respectively, and the colors represent various outer shell radii used to pick out the upstream magnetic field. Each color represents a different value of the outer radius of the spherical shell. Evidently, the upstream field distribution does not change much even after using larger radii for the spherical shells. Note that $b^{r}$, $b^{\theta}$ and $b^{\phi}$ do not all have the same units.  }
    \label{fig:bcorner_plot}
\end{figure}

The initial wavevectors of the geodesics depend on the upstream magnetic field when the effect of beaming is included in the model. We pick the upstream magnetic field vector by scanning a small spherical shell around every cell in the current sheet and then picking the magnetic field vector in the cube with the largest fluid frame magnitude. Because the reconnecting field is less ordered than the upstream field, we set the inner radius of the shell to be $0.25$ $r_{\rm{g}}$ to not pick any magnetic field vector within the current sheet and vary the outer radius of the shell. Here, we compare the distributions of the selected upstream magnetic field using different values of the outer shell radii and show that the distribution of the identified upstream field does not depend strongly upon the choice of the outer radius. Figure  \ref{fig:bcorner_plot} is a plot of $2$D histograms of the components of the upstream field as a function of $r$ using different outer radii for the shells. The blue ($R_{\rm out}=0.4$ $r_{\rm{g}}$), red ($R_{\rm out}=0.35$ $r_{\rm g}$) and orange ($R_{\rm out}=0.3$ $r_{\rm{g}}$)  distributions are similar, demonstrating that the selected upstream is independent of the size of the shell. 
\section{Convergence of Ray tracing} \label{Appendix: Convergence of the Lightcurve}
\begin{figure}
    \centering
    \includegraphics[scale=0.6]{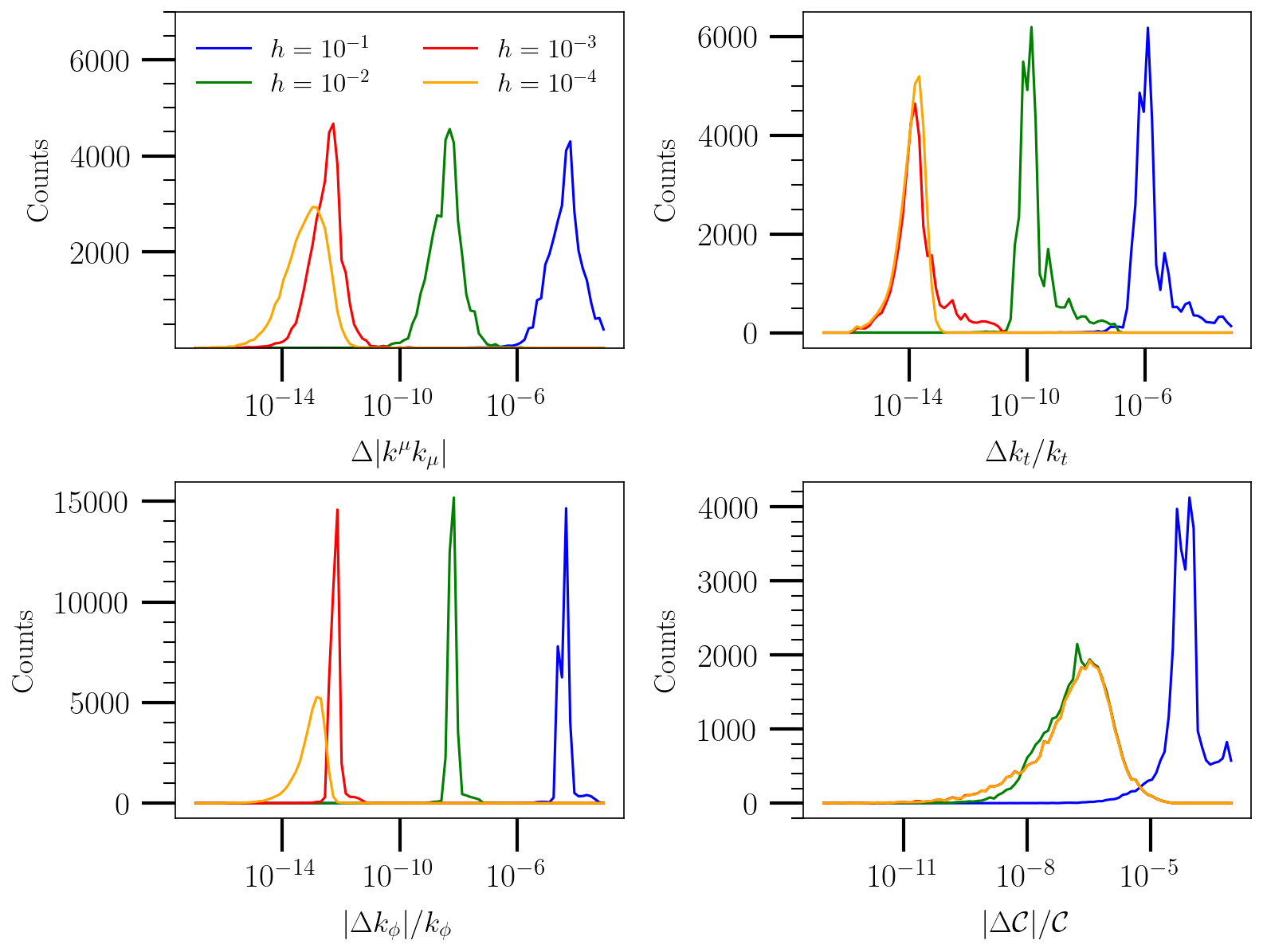}
    \caption{The relative error in the norm, energy, angular momentum and Carter's constant of the $60000$ superphotons between the start and end of the numerical integration. }
    \label{fig:convergence_constants}
\end{figure}

There are various methods to test the accuracy and convergence of ray tracing method to compute the lightcurves. One can check whether the constants of motion (norm of the wavevector, energy, angular momentum and Carter's constant) are conserved along a select few geodesics. However, this does not necessarily guarantee convergence of the entire lightcurve. There may also be the issue of under-sampling, i.e., a very small number of superphotons make it to certain bins where the lightcurve is being computed. The lightcurve must also be numerically converged such that changing the stepsize of the integrator does not change the result. Here, we test our lightcurves using all of the above-mentioned metrics.
\newline
We integrate an ensemble of $60000$ superphotons randomly initialized to lie in the current sheet with a wavevector constructed as described in Section \ref{Methods}. We use a RK4 integrator for the geodesic integration with adaptive timesteping taken from \cite{GRMONTY_2009ApJS..184..387D}. We perform a scan over various initial timesteps and measure the relative change in the norm, energy, angular momentum and Carter's constant evaluated at the initial (in the current sheet) and final (at $10^4 r_{\rm{g}}$) positions of the superphotons. The superphotons have an initial wavevector norm $k^{\mu}k_{\mu} = 0$, and therefore we take the final value of $|k^{\mu}k_{\mu}|$ as the change in norm. The energy and angular momentum are calculated as $E = g_{\alpha t} k^{\alpha}$ and $L = g_{\alpha \phi} k^{\alpha}$ respectively, and we report $|(E_f - E_i)| / (E_i + \epsilon)$ and $|(L_f - L_i)| / (L_i + \epsilon)$, where $\epsilon = 10^{-20}$. Finally, we also compute the relative change in the Carter's constant for the superphotons, which requires a coordinate transformation to BL coordinates. As is evident from Figure \ref{fig:convergence_constants}, the norm and the relative change in angular momentum are conserved to $1$ part in $10^{-13}$, the energy to roughly $1$ part in $10^{-14}$ and the Carter's constant to $1$ part in $10^{-6}$. The relatively large difference in the latter is partly due to errors in the coordinate transformation involving a non-linear root find.

The numerical convergence of the arrival times of the superphotons is tested next. We once again integrate $60000$ randomly selected geodesics using initial step sizes of $10^{-1}, 10^{-2}, 10^{-3} 10^{-4} \text{ and }  10^{-5}$ $t_{\rm g}$ with a RK-4 integrator and adaptive timestepping. We construct the lightcurve by binning the $t$ coordinates of the superphotons once they reach $10^4 r_{\rm{g}}$. The distributions converge to $\Delta t \lesssim 1 r_{\mathrm{g}}/c$ for $h=10^{-4} $ and $10^{-5}$ $t_{\rm g}$, despite the relatively small uncertainties in the constants of motion even at larger stepsizes. This is because the steps are taken in $\log r$, and convergence in $t$ is desired.  Therefore, we use an initial stepsize of $h = 10^{-4}$ for our geodesic integration presented in this paper.

\begin{figure}
    \centering
    \includegraphics[scale=0.5]{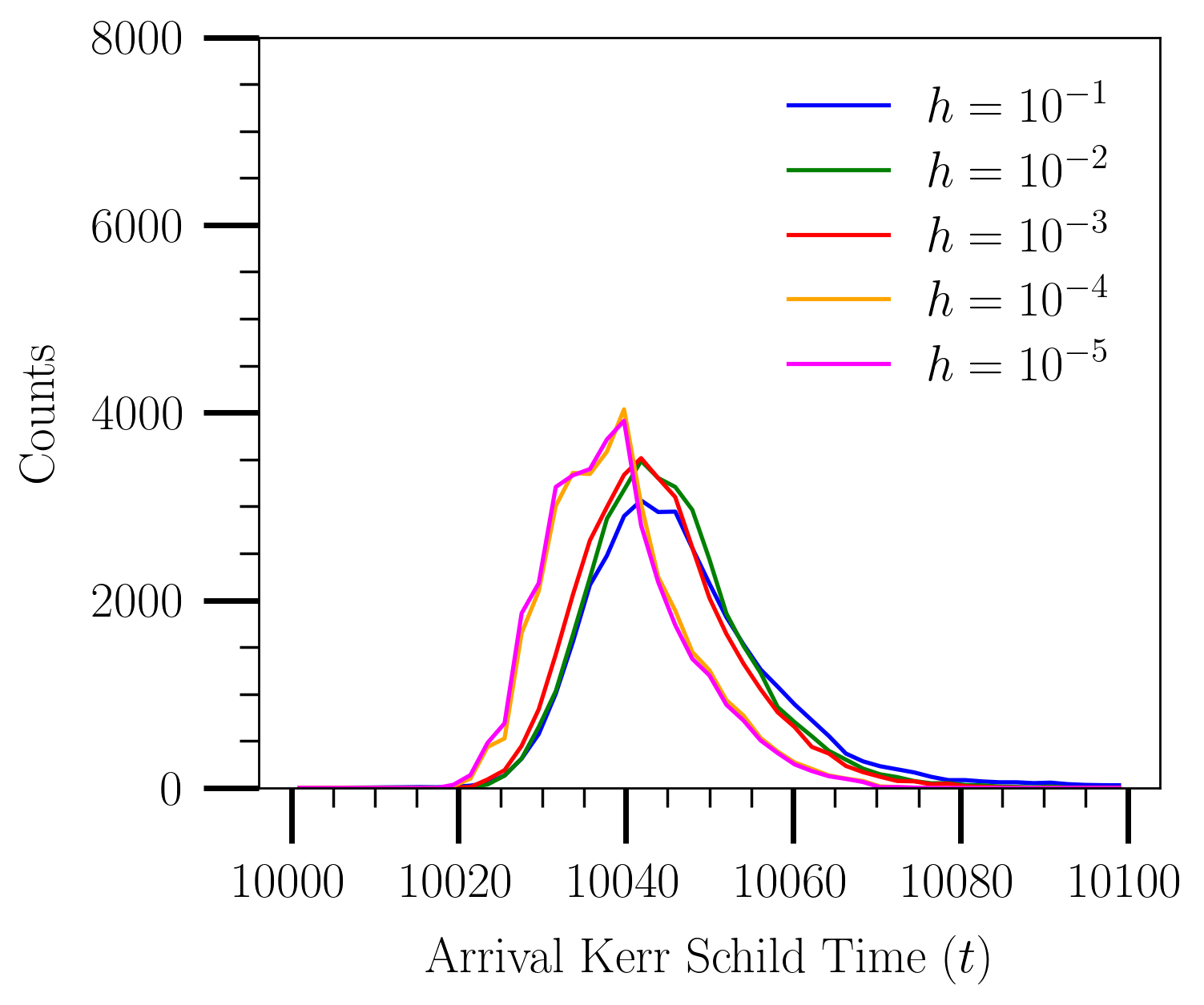}
    \caption{Convergence test by integrating the geodesics from a single simulation snapshot using different initial stepsizes. This is done by binning the $t$ coordinate of superphotons that reach $10^4 r_{\rm{g}}$. We see acceptable convergence for an initial stepsize of $ h \leq 10^{-4}$ $t_{\rm g}$, which is what we use for our geodesic integration presented in this paper.}
    \label{fig:stepsize-convergence}
\end{figure}

Finally, we test the lightcurve convergence by sampling a different number of superphotons per point. That is, we initialize geodesics such that $N_i = c_i^{1/2}, 5c_{i}^{1/2}$ and $10c_i^{1/2}$ such that each point in the current sheet has at least $1, 5$ and $10$ superphotons initialized respectively. The arrival $t, \theta$ and $\phi$ coordinate distributions of the weighted superphotons are almost identical. We, therefore, use $N_i=c_i^{1/2}$ (i.e., at least $1$ superphoton per point) for our ray traced lightcurves. 

\begin{figure}
    \centering
    \includegraphics[scale=0.5]{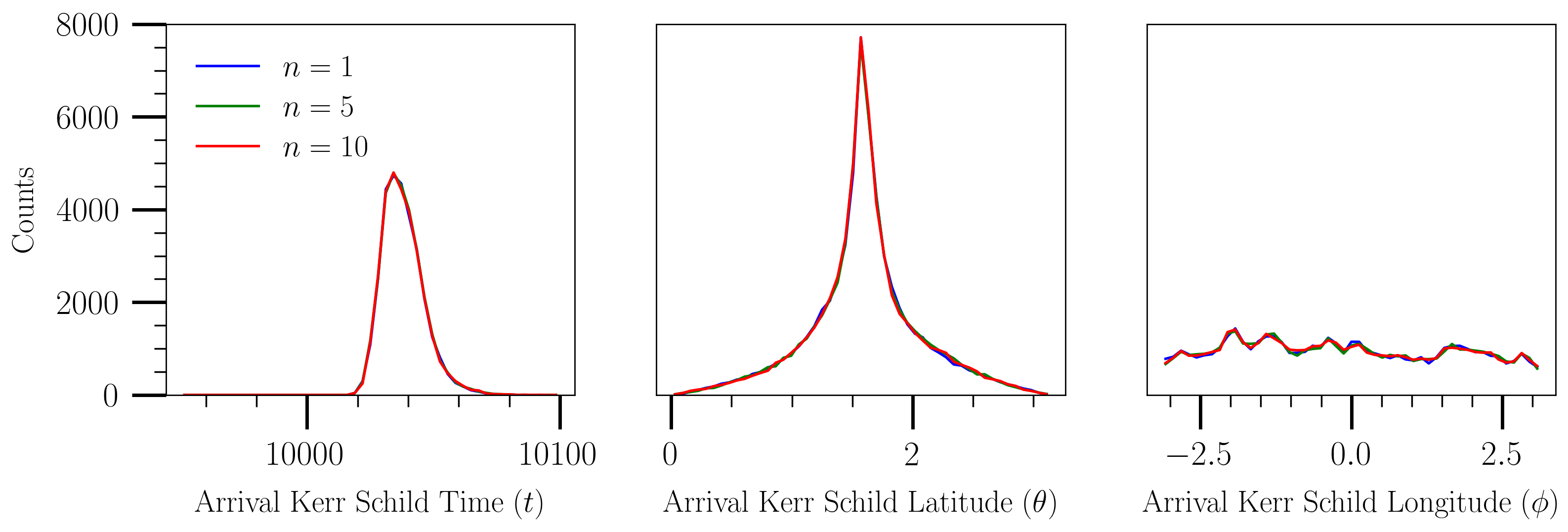}
    \caption{We test the convergence of the lightcurve by using a different number of minimum superphotons from each point in the current sheet. Here, we test the convergence in the $t,\theta$, and $\phi$ coordinates by launching at least $1, 5$ or $10$ superphotons per cell in the current sheet. The coordinates of superphotons are binned once they cross $10^4 r_{\textrm{g}}$ after being weighted by the process described in Section \ref{subsec: photon initialization}. The final coordinate distributions for $N=1, 5$ and $10$ lie on top of each other, indicating that superphotons are adequately sampled.}
    \label{fig:num-geodesics-convergence}
\end{figure}

\end{document}